\documentclass[review]{elsarticle}

\usepackage{graphicx}
\usepackage{dcolumn}
\usepackage{bm}
\usepackage{epsfig}
\usepackage{booktabs}
\usepackage{subfigure}
\usepackage{graphics}
\usepackage{amssymb}
\usepackage{amsmath}
\usepackage{array}
\usepackage{color}
\usepackage{multirow}
\graphicspath{{./pic/}}
\journal{Computers \& Mathematics with Applications}

\begin{document}

\begin{frontmatter}

\title{A finite-difference lattice Boltzmann model with second-order accuracy of time and space for incompressible flow}

\author[mymainaddress]{Xinmeng Chen}
\author[mymainaddress,mysecondaryaddress]{Zhenhua Chai}
\author[mymainaddress,mythirdaddress]{Huili Wang}
\author[mymainaddress,mysecondaryaddress]{Baochang Shi\corref{mycorrespondingauthor}}
\cortext[mycorrespondingauthor]{Corresponding author}
\ead{shibc@hust.edu.cn}
\address[mymainaddress]{School of Mathematics and Statistics, Huazhong University of Science and Technology, Wuhan 430074, China}
\address[mysecondaryaddress]{Hubei Key Laboratory of Engineering Modeling and Scientific Computing, Huazhong University of Science and Technology, Wuhan 430074, China}
\address[mythirdaddress]{School of Mathematics and Computer Science, Wuhan Textile University, Wuhan, 430073, China}

\begin{abstract}
  In this paper, a kind of finite-difference lattice Boltzmann method with
  the second-order accuracy of time and space (T2S2-FDLBM) is proposed. In
  this method, a new simplified two-stage fourth order time-accurate
  discretization approach is applied to construct time marching scheme,
  and the spatial gradient operator is discretized by a mixed difference
  scheme to maintain a second-order accuracy both in time and space. It is
  shown that the previous finite-difference lattice Boltzmann method
  (FDLBM) proposed by Guo \cite{guo2003explicit} is a special case of the
  T2S2-FDLBM. Through the von Neumann analysis, the stability of the
  method is analyzed and two specific T2S2-FDLBMs are discussed. The two
  T2S2-FDLBMs are applied to simulate some incompressible flows with the
  non-uniform grids. Compared with the previous FDLBM and SLBM, the
  T2S2-FDLBM is more accurate and more stable. The value of the
  Courant-Friedrichs-Lewy condition number in our method can be up to 0.9,
  which also significantly improves the computational efficiency.
\end{abstract}


\end{frontmatter}


\section{\label{sec:level1}Introduction}
In the past 30 years, the lattice Boltzmann method (LBM) has received
increasing attention, and also made great progress in the fields of
microscale flows \cite{raabe2004overview,guo2013lattice,CY2002application},
porous media flows \cite{guo2002lattice,chai2019lattice}, multiphase flows
\cite{shan1993Lattice,Shan1994simulation,swift1995Lattice,wang2016comparative,chai2018comparative,Wang2019a}
and turbulent flows \cite{chen2003extended,strumolo1997new}. With the
approach of the interpolation \cite{he1996some,yu2002multi}, LBM can be
implemented on nonuniform grids. In the LBM, the refinement scheme of
non-uniform mesh mainly includes static mesh refinement
\cite{filippova1998grid,lin2000lattice,lagrava2012advances} and dynamic mesh
refinement
\cite{crouse2003lb,tolke2006adaptive,wu2011solution,chen2011lattice,fakhari2014finite}.
In addition, there are several methods combined LBM with finite-difference
\cite{wang2017Finite,Xiu2018general,lee2003eulerian,fakhari2014finite,eitel2013lattice},
finite-volume \cite{Guo2013DUGKS,wu2016DUGKS}, finite-element approaches
\cite{lee2001characteristic} and so on. These methods can promote the
geometrical flexibility of the LBM. Unlike SLBM, the discrete-velocity in
these methods is decoupled with lattice and time steps, and thus, the
non-uniform meshes can be used to improve computational efficiency and
accuracy of LBM.

In 1995, Reider and Sterling first proposed a finite-difference lattice
Boltzmann equation (FDLBE) for the simulation of the incompressible
Navier-Stokes equations \cite{reider1995accuracy}, then Cao and Chen
examined more details of FDLBE \cite{cao1997physical}. Mei and Shyy
developed the FDLBM on curvilinear coordinates \cite{mei1998finite}. Based
on this work, Guo et al. proposed an implicit treatment for collision term
of the Bhatnagar-Gross-Krook Boltzmann equation, and presented a mixed
difference scheme to discrete the advection term \cite{guo2003explicit}. In
addition, Guo et al. developed a new FDLBM for dense binary mixtures, where
the second-order Lax Wendroff scheme and first-order Euler's formula are
used to discrete space and time derivatives, respectively
\cite{guo2005finite}. Wang et al. proposed a high-order FDLBM to deal with
the compressibility and non-linear shock wave effects in the resonator, in
which a third-order implicit-explicit Runge-Kutta scheme and a fifth-order
weighted essentially non-oscillatory (WENO) scheme are used for time and
space discretization \cite{wang2009implicit}. In 2013, Amin and Sun studied
the stability condition of the FDLBM \cite{el2013stability}. Subsequently,
Kim and Yang incorporates the immersed boundary method into the FDLBM
\cite{kim2016immersed}. Besides, the multi-speed model was also combined
with the FDLBM \cite{watari2009velocity}, and the work on FDLBM for
three-dimensional incompressible flows \cite{ezzatneshan2019simulation} was
also conducted.

Up to now, most of high-order FDLBMs are implemented by using Runge-Kutta
scheme for time discretization, and fifth-order WENO scheme or fourth-order
compact finite-difference scheme for space discretization
\cite{wang2009implicit,hejranfar2014high,hejranfar2015simulation,hejranfar2018preconditioned,ezzatneshan2019simulation}.
The high-order FDLBMs can be used to improve the accuracy and convergent
speed, but those methods are at the expense of overcomplicated calculation,
which loses the simplicity of SLBM. Moreover, most of the current work on
high order FDLBM are implemented on uniform grid which neglects the
computational efficiency brought by non-uniform grids. In addition, the
simple FDLBM proposed by Guo \cite{guo2003explicit} retains the
computational framework of SLBM, but the computational efficiency was not
improved greatly. This limitation results from the value of
Courant-Friedrichs-Lewy (CFL) condition number. In the implementation of
FDLBM, CFL condition number is proportional to the time step. The
theoretical range of CFL condition number is $(0,1)$, but it is usually
around 0.1 in the numerical simulations. In contrast, the value of the CFL
condition number is 1 in the SLBM. Compared with SLBM, the application of
non-uniform grid in FDLBM may bring higher computational efficiency and
accuracy, but the smaller value of CFL condition number will also influence
the efficiency.

In order to solve this problem, a new FDLBM with a second-order accuracy
both in space and time is proposed. To simplify following discussion, the
FDLBM developed by Guo \cite{guo2003explicit} is marked as T1S2-FDLBM, and
the present FDLBM is denoted by T2S2-FDLBM. The theoretical analysis and
numerical result show that T2S2-FDLBM has a second-order accuracy in space
and time, and also, the value of the CFL condition number in T2S2-FDLBM can
be increased up to 0.9. At the same time, with the application of
non-uniform grid, the T2S2-FDLBM is more efficient and more accurate than
SLBM and T1S2-FDLBM.

The rest of the paper is organized as follows. In Sec.~\ref{sec:level2}, the
T2S2-FDLBM is constructed from time and spatial discretization. We also
analyze the stability of the method by Von Neumann analysis. Then some
numerical simulation are conducted in Sec.~\ref{sec:level3}, and finally,
some conclusions are given in Sec.~\ref{sec:level4}.

\section{\label{sec:level2}Numerical methods}
In this section, the Boltzmann equation will be discreted by some different
schemes. Inspired by Wu et al. \cite{wu2018third}, we use a new simplified
two-stage fourth order time-accurate discretization (TFTD) method to
construct the time marching scheme. And similar to the spatial
discretization in \cite{guo2003explicit}, the gradient term is discreted by
a mixed difference scheme which incorporates the central difference and the
second-order upwind-difference schemes. Moreover, the Von Neumann analysis
would be applied to evaluate the numerical stability of the T2S2-FDLBM and
determine the model parameters.

\textbf{A. Time discretization}

The time-dependent Boltzmann equation with a force term can be written
follows
\begin{equation}
\frac{{\partial f}}{\partial t} = L(f)+\bar{\Omega} (f):=M,
\label{eq:2.1}
\end{equation}
 where \emph{f} is particle distribution function. The gradient term \emph{L} and $\bar{\Omega}$ can be expressed as
\begin{equation}
\ L(f)=-\bm{\xi} \cdot \nabla f,
\label{eq:2.2}
\end{equation}
\begin{equation}
\ \bar{\Omega}(f)=\Omega(f)+F,
\label{eq:2.3}
\end{equation}
 where $\bm{\xi}$ represents the particle velocity and the $\bar{\Omega}$ contains the force term \emph{F} and
collision term $\Omega$. Integrating Eq.~\eqref{eq:2.1} over the time
interval $[t_n,t_n+\Delta t]$ yields
\begin{equation}
\ f^{n+1}=f^{n}+\int_{t_n}^{t_n+\Delta t} M[f(t)]dt,
\label{eq:2.5}
\end{equation}
where $f^n=f(\bm{x},\bm{\xi},t_n)$. According to the previous work
\cite{li2016two}, the chain rule and Cauchy-Kovalevskaya theorem can be used
to deal with time derivative of $M(f)$ at $t=t_n$,
\begin{equation}
\ \frac{\partial}{\partial t}M(f^n)=M_f(f^n)M(f^n),
\label{eq:2.6}
\end{equation}
where $M_f=dM/df$. To ensure the second-order accuracy in time, applying
Taylor expansion to $M(f)$, we have
\begin{equation}
\begin{split}
\ M(f) & =  M(f^n)+(t-t_n)\frac{\partial}{\partial t}M(f^n)+O(\triangle t^2)\\
\ & = M(f^n)+(t-t_n)M_f(f^n)M(f^n)+O(\triangle t^2).
\end{split}
\label{eq:2.7}
\end{equation}
Consequently, the time integral term in Eq.~\eqref{eq:2.5} can be
approximated as
\begin{equation}
\ \int_{t_n}^{t_n+\Delta t}M(f)dt =\Delta t M(f^n) +\frac{\Delta t^2}{2}M_f(f^n)M(f^n)+O(\Delta t^3).
\label{eq:2.8}
\end{equation}
Then, to construct a numerical scheme to Eq.~\eqref{eq:2.5}, we introduce
the intermediate variable $f^*=f(\bm{x},\bm{\xi},t_*)$ at time
$t_*=t_n+A\Delta t$. Using Taylor series analysis to the intermediate
variable, we obtain
\begin{equation}
\ f^*=f^n+A \Delta t M(f^n)+\frac{1}{2} A^2 \Delta t^2\frac{\partial}{\partial t}M(f^n)+O(\Delta t^3).
\label{eq:2.9}
\end{equation}
From Eq.~\eqref{eq:2.5}, one can also have
\begin{eqnarray}
\  f^{n+1} & = & f^n+\Delta t \left[ B_0 M(f^n)+B_1 M(f^*)+B_2M(f^{n+1}) \right]\label{eq:2.10a},\\
\  f^{n+1} & = & f^n+ \Delta t \left[ B_0 L(f^n)+B_1 L(f^*)+B_2L(f^{n+1}) \right]\nonumber\\
\          &   & +\Delta t \left[ B_0 \bar{\Omega}(f^n)+B_1 \bar{\Omega}(f^*)+B_2 \bar{\Omega}(f^{n+1}) \right]\label{eq:2.10b},
\end{eqnarray}
where $B_0$, $B_1$, $B_2$ and $A$ are adjustable parameters. By expanding
the $f^*$ and $f^{n+1}$ at $f^n$, we have
\begin{eqnarray}
\ \Delta t \left[ B_0 M(f^n)+B_1 M(f^*)+B_2M(f^{n+1}) \right] = \Delta t \left( B_0+B_1+B_2 \right )M(f^n)& & \nonumber   \\
                                                            +\frac{\Delta t^2}{2} \left[2 (AB_1+B_2)\right ]M_f(f^n)M(f^n)+O(\Delta t^3).&   &
\label{eq:2.11}
\end{eqnarray}
Through a comparison of Eqs.~\eqref{eq:2.8} and \eqref{eq:2.11}, the
following relations can be derived,
\begin{subequations}
\begin{equation}
\ B_0+B_1+B_2=1,
\label{eq:2.12a}
\end{equation}
\begin{equation}
\ AB_1+B_2=\frac{1}{2},
\label{eq:2.12b}
\end{equation}
\end{subequations}
where $0\leq B_0,B_1,A\leq 1$ and $0\leq B_2\leq 1/2$. Here it should be
noted that the number of equations is less than that of parameters, and two
of those parameters can be adjusted flexibly. Besides, it is clear that
Eq.~\eqref{eq:2.10a} is a first-order scheme when only Eq.~\eqref{eq:2.12a}
is satisfied, while it would be a second-order scheme when both
Eqs.~\eqref{eq:2.12a} and \eqref{eq:2.12b} are satisfied.

 To generalize the method, the coefficients in $L(f)$ and $\bar{\Omega}(f)$ can be
designed as follows,
\begin{eqnarray}
\  f^{n+1} & = & f^n +\Delta t \left[ \tilde{B_0} L(f^n)+\tilde{B_1} L(f^*)+\tilde{B_2}L(f^{n+1}) \right]\nonumber\\
           &   &+\Delta t \left[ \bar{B_0} \bar{\Omega}(f^n)+\bar{B_1} \bar{\Omega}(f^*)+\bar{B_2} \bar{\Omega}(f^{n+1}) \right].
\label{eq:2.13}
\end{eqnarray}
Equation \eqref{eq:2.10b} can be considered as a special case of
Eq.~\eqref{eq:2.13} when the parameters satisfy the relations,
$\tilde{B_0}=\bar{B_0},\tilde{B_1}=\bar{B_1},\tilde{B_2}=\bar{B_2}.$
Similarly, after a comparison of Eqs. \eqref{eq:2.9} and \eqref{eq:2.13}, we
have
\begin{subequations}
\begin{equation}
\tilde{B_0}+\tilde{B_1}+\tilde{B_2} = 1,   \bar{B_0}+\bar{B_1}+\bar{B_2}  =  1,
\label{eq:2.14a}
\end{equation}
\begin{equation}
 A\tilde{B_1}+\tilde{B_2} = \frac{1}{2},A\bar{B_1}+\bar{B_2}  =  \frac{1}{2},
\label{eq:2.14b}
\end{equation}
\end{subequations}
where $0\leq \tilde{B_0},\tilde{B_1},A\leq 1$, $0\leq \tilde{B_2}\leq 1/2$
and $0\leq \bar{B_0},\bar{B_1},A\leq 1$, $0\leq \bar{B_2}\leq 1/2$.

 \textbf{Remark I}. We noted that T1S2-FDLBM is a special case of
 Eq.~\eqref{eq:2.13}. Actually, according to Eq.~\eqref{eq:2.14a}, the T1S2-FDLBM in Ref \cite{guo2003explicit} can be obtained when the
following relations are satisfied,
\begin{equation}
A=0,\tilde{B_0}=1,\tilde{B_1}=\tilde{B_2}=0,\bar{B_0}=\frac{1}{2},\bar{B_1}=0,\bar{B_2}=\frac{1}{2}.
\label{eq:2.16}
\end{equation}
The evolution equation of T1S2-FDLBM reads
\begin{equation}
 f^{n+1}  =  f^n+\Delta t L(f^n)+\frac{1}{2}\Delta t \left[\bar{\Omega}(f^n)+\bar{\Omega}(f^{n+1}) \right]+O(\Delta t^2).
 \label{eq:2.17}
\end{equation}
We would like to point out that Eq.~\eqref{eq:2.17} can be rewritten in an
explicit form,
\begin{equation}
 \tilde{f}^{n+1}  = \tilde{f} ^{+,n}+\Delta t  L(f^n)+O(\Delta t^2),
 \label{eq:2.18}
\end{equation}
where
\begin{equation}
 \tilde{f}  = f-\frac{1}{2}\Delta t \bar{\Omega},
 \label{eq:2.19}
\end{equation}
\begin{equation}
 \tilde{f}^+  =f+\frac{1}{2}\Delta t \bar{\Omega}= \frac{2\tau -\Delta t}{2\tau+\Delta t}\tilde{f}+
 \frac{2\Delta t}{2\tau+\Delta t}f^{eq}+\frac{2\tau \Delta t}{2\tau+\Delta t}F.
\label{eq:2.20}
\end{equation}

In order to construct a second-order time marching scheme of FDLBM, there
are two sets of parameters need to be considered (we refer the reader to
section C for details). The first one is designed as
\begin{equation}
A=\frac{1}{2}, \tilde{B_0}=0, \tilde{B_1}=1, \tilde{B_2}=0, \bar{B_0}=\frac{1}{2}, \bar{B_1}=0, \bar{B_2}=\frac{1}{2},
\label{eq:2.21}
\end{equation}
and the corresponding evolution equation can be rewritten as
\begin{equation}
 f^{n+1}  =  f^n+\Delta t  L(f^*)+\frac{1}{2} \Delta t \left[\bar{\Omega}(f^n)+\bar{\Omega}(f^{n+1}) \right]+O(\Delta t^2).
\label{eq:2.22}
\end{equation}
In addition, we can also determine the other one as
\begin{equation}
\tilde{B_0}=\frac{1}{2}, \tilde{B_1}=0, \tilde{B_2}=\frac{1}{2}, \bar{B_0}=\frac{1}{2},\bar{B_1}=0, \bar{B_2}=\frac{1}{2},\quad for\quad A\in[0,1],
\label{eq:5.21}
\end{equation}
and the evolution equation can also be obtain,
\begin{equation}
 f^{n+1}  =  f^n+\frac{\Delta t}{2} ( L(f^n)+L(f^{n+1}))+\frac{1}{2} \Delta t \left[\bar{\Omega}(f^n)+\bar{\Omega}(f^{n+1}) \right]+O(\Delta t^2).
\label{eq:5.22}
\end{equation}
Similarly to Eq.~\eqref{eq:2.17}, Eqs.~\eqref{eq:2.22} and \eqref{eq:5.22}
can also be written in the explicit forms,
\begin{equation}
 \tilde{f}^{n+1}  = \tilde{f} ^{+,n}+\Delta t  L(f^*)+O(\Delta t^2),
\label{eq:2.23}
\end{equation}
\begin{equation}
 \tilde{f}^{n+1}  = \tilde{f} ^{+,n}+\frac{\Delta t}{2} ( L(f^n)+L(f^{n+1}))+O(\Delta t^2).
\label{eq:5.23}
\end{equation}

To make a distinction between two T2S2-FDLBMs, the first one described by
Eq.~\eqref{eq:2.23} is denoted by T2S2-FDLBM1, while the second one given by
Eq.~\eqref{eq:5.23} is marked as T2S2-FDLBM2.

\textbf{B. Space discretization}

Here the method of integrating along the characteristic line is used to
calculate the distribution function at intermediate moment. The discrete
form of Eq.~\eqref{eq:2.1} can be expressed as
\begin{equation}
 \tilde{f}^{n+1}_i  = \tilde{f} ^{+,n}_i+\Delta t  L(f_i^*),
\label{eq:2.26}
\end{equation}
or
\begin{equation}
 \tilde{f}^{n+1}_i  = \tilde{f} ^{+,n}_i+\frac{\Delta t}{2} ( L(f_i^n)+L(f_i^{n+1})),
\label{eq:5.26}
\end{equation}
where $f_i^n=f(\bm{x},\xi_i,t_n)$, and the gradient terms can be rewritten
as
\begin{equation}
  L(f_i)=-\bm{\xi}_i\cdot \nabla f_i=-\xi_{i\alpha}\frac{\partial f_i}{\partial \chi_\alpha},
\label{eq:2.27}
\end{equation}
where $f_i$ represents the distribution function $f^n_i$, $f^{n+1}_i$ or
$f^*_i$. Therefore, the evaluation of the distribution function $f^*_i$ or
$f^{n+1}_i$ is the key to calculate the gradient term. Considering the
following Boltzmann equation,
\begin{equation}
 \frac{\partial f_i}{\partial t}+\bm{\xi}_i \cdot \nabla f_i=\bar{\Omega_i},
\label{eq:2.28}
\end{equation}
and integrating Eq.~\eqref{eq:2.28} along the characteristic line $x+\xi_i
t$ over $[0,h]$, we have
\begin{equation}
 f(\bm{x}+\bm{\xi}_i t,\xi_i,t)-f(\bm{x},\bm{\xi}_i,t)=\int^h_0 \bar{\Omega}(\bm{x}+\bm{\xi}_i s,\bm{\xi}_i,s)ds,
\label{eq:2.29}
\end{equation}
where $h$ is the time step, $h=\Delta t/2$ in the T2S2-FDLBM1, and $h=\Delta
t$ in the T2S2-FDLBM2.

When the trapezoidal formula is applied to approximate the integral of the
collision term in Eq.~\eqref{eq:2.29}, we can obtain
\begin{equation}
 f(\bm{x},\bm{\xi}_i,t+h)-f(\bm{x}-\bm{\xi}_i h,\bm{\xi}_i,t)=\frac{h}{2}\left[ \bar{\Omega}(\bm{x},\bm{\xi}_i,t+h)+\bar{\Omega}(\bm{x}-\bm{\xi}_i h,\bm{\xi}_i,t) \right].
\label{eq:2.30}
\end{equation}
Introducing a new variable $\bar{f}_i$,
\begin{equation}
 \bar{f}_i=f_i-\frac{h}{2}\bar{\Omega}_i,
\label{eq:2.31}
\end{equation}
or equivalently,
\begin{equation}
 f_i=\frac{2\tau}{2\tau+h}\bar{f}_i+\frac{h}{2\tau+h}f_i^{eq}+\frac{h\tau}{2\tau+h}F_i,
\label{eq:2.32}
\end{equation}
we can rewrite Eq.~\eqref{eq:2.30} as
\begin{equation}
 \bar{f}(\bm{x},\bm{\xi}_i,t+h)=\bar{f}^{+,h}(\bm{x}-\bm{\xi}_i h,\bm{\xi}_i,t),
\label{eq:2.33}
\end{equation}
where
\begin{equation}
 \bar{f}_i^{+,h}=f_i+\frac{h}{2}\bar{\Omega}_i=\frac{2\tau-h}{2\tau+h}\bar{f}_i+\frac{2h}{2\tau+h}f_i^{eq}+\frac{2h\tau}{2\tau+h}F_i.
\label{eq:2.34}
\end{equation}

To ensure that the Eq.~\eqref{eq:2.26} achieves a second-order accuracy, the
term $ L(f_i^*) $ should have a first-order accuracy. With the Taylor
expansion, we can express $\bar{f}^{+,h}(\bm{x}-\bm{\xi}_i h,\bm{\xi}_i,t)$
as
\begin{equation}
 \bar{f}^{+,h}(\bm{x}-\bm{\xi}_ih,\bm{\xi}_i,t)=\bar{f}^{+,h}(\bm{x},\bm{\xi}_i,t)-h\xi_i \nabla\cdot \bar{f}^{+,h}(\bm{x},\bm{\xi}_i,t).
\label{eq:2.35}
\end{equation}
In order to simplify the calculation, we use the same difference scheme to
deal with the gradient terms in Eqs. \eqref{eq:2.26}, \eqref{eq:5.26} and
\eqref{eq:2.35}. Usually, the gradient term can be approximated by the
central difference or upwind-difference schemes. However, the second-order
upwind-difference scheme is more stable and the central-difference scheme
has smaller numerical dispersion for high Reynolds number problems. For this
reason, a mixed-difference scheme which combines the central-difference and
second-order upwind difference schemes is adopted here,
\begin{equation}
 \frac{\partial f_i}{\partial \chi_\alpha} \Bigg{|}_m=\eta \frac{\partial f_i}{\partial \chi_\alpha} \Bigg{|}_c+(1-\eta)\frac{\partial f_i}{\partial \chi_\alpha} \Bigg{|}_u ,
\label{eq:2.36}
\end{equation}
where $f_i$ represents $f_i^*$ or $\bar{f}_i^{+,h}$, and the parameter $\eta
\in [0,1]$. The terms $\dfrac{\partial f_i}{\partial \chi_\alpha}
\Bigg{|}_u$ and $\dfrac{\partial f_i}{\partial \chi_\alpha} \Bigg{|}_c$
represent second up-wind difference and central-difference schemes, and they
are defined as
\begin{equation}
 \frac{\partial f_i}{\partial \chi_\alpha} \Bigg{|}_c=\frac{f_i(\chi_\alpha+\Delta \chi_\alpha,t)-f_i(\chi_\alpha-\Delta \chi_\alpha,t)}{2\Delta \chi_\alpha},
\label{eq:2.37}
\end{equation}
\begin{equation}
 \frac{\partial f_i}{\partial \chi_\alpha} \Bigg{|}_c=
 \begin{cases}
 \dfrac{3f_i(\chi_\alpha,t)-4f_i(\chi_\alpha-\Delta \chi_\alpha,t)+f_i(\chi_\alpha-2\Delta \chi_\alpha,t)}{2\Delta \chi_\alpha}, \quad &if \quad {c_{i\alpha} \geq 0},\\
 -\dfrac{3f_i(\chi_\alpha,t)-4f_i(\chi_\alpha+\Delta \chi_\alpha,t)+f_i(\chi_\alpha+\Delta \chi_\alpha,t)}{2\Delta \chi_\alpha}, \quad &if \quad {c_{i\alpha}< 0}.
 \end{cases}
 \label{eq:2.38}
\end{equation}

\textbf{C. Analysis of the T2S2-FDLBM}

In this part, the von Neumann method is used to analyzed the numerical
stability of the T2S2-FDLBM, and the force term is ignored to simplify the
analysis. The evolution equation \eqref{eq:2.13} can be rewritten as
\begin{equation}
\begin{split}
\  f^{n+1}_i  &=  f^n_i +\Delta t \left[ \tilde{B_0} L(f^n_i)+\tilde{B_1} L(f^*_i) +\tilde{B_2} L(f^{n+1}_i)\right]\\
            &  + \Delta t \left[ \bar{B_0}\Omega(f^n_i)+\bar{B_1}\Omega(f^*_i)+\bar{B_2}\Omega(f^{n+1}_i) \right].
\label{eq:4.1}
\end{split}
\end{equation}
According to the fact
$f(\bm{x},\bm{\xi}_i,t+h)=f(\bm{x},\bm{\xi}_i,t)+h\partial_tf^{eq}(\bm{x},\bm{\xi}_i,t)+O(h^2)$
\cite{guo2013discrete}, expanding $f_i^*$ in Eq.~\eqref{eq:4.1} yields
\begin{equation}
\begin{split}
\ & f^{n+1}_i-\Delta t\bar{B_2}\Omega(f^{n+1}_i)-\Delta t\tilde{B_2}L(f^{n+1}_i) =  f^n_i+\Delta t(1-\bar{B_2})\Omega(f^{n}_i)\\
   &   +\Delta t(1-\tilde{B_2})L(f^{n}_i)+\Delta t(\frac{1}{2}-\tilde{B_2})L(\Delta t\partial_tf_i^{eq,n})-\omega(\frac{1}{2}-\bar{B_2})\Delta t\partial_tf_i^{eq,n}.
\label{eq:4.2}
\end{split}
\end{equation}
where $f_i^{eq,n}=f^{eq}(\bm{x},\bm{\xi}_i,t_n)$. Then if the Euler formula
is used to deal with time derivative, one can obtain
\begin{equation}
\begin{split}
\  &f^{n+1}_i-\Delta t\bar{B_2}\Omega(f^{n+1}_i)+\omega(\frac{1}{2}-\bar{B_2})f_i^{eq,n+1}+\Delta t\tilde{B_2}\xi\cdot \nabla f^{n+1}_i= \\
  &-\Delta t(\frac{1}{2}-\tilde{B_2})\xi\cdot \nabla f^{eq,n+1}_i+f^n_i+\Delta t(1-\bar{B_2})\Omega(f^{n}_i)-\Delta t(1-\tilde{B_2})\xi\cdot \nabla f^{n}_i\\
  &+\Delta t(\frac{1}{2}-\tilde{B_2})\xi\cdot \nabla f^{eq,n}_i+\omega(\frac{1}{2}-\bar{B_2})f_i^{eq,n},
\label{eq:4.3}
\end{split}
\end{equation}
where $\omega=\Delta t/\tau$. To conduct a linear stability analysis, $f_i$
is expanded as
\begin{equation}
f_i(\bm x,t)=\overline{f_i^{eq}(\bm x,t)}+f'_i(\bm x,t),
\label{eq:4.4}
\end{equation}
where $\overline{f_i^{eq}(\bm x,t)}$ represents the global equilibrium
distribution. It only depends on the mean value of density $\rho$ and
velocity $u$, and does not vary with time and space. $f'_i$ is the
fluctuating quantity of $f_i$. With the help of Eq. \eqref{eq:4.4},
Eq.~\eqref{eq:4.3} can be written as
\begin{equation}
\begin{split}
[(1+\omega\bar{B_2})\delta_{ij}+\omega(\frac{1}{2}-2\bar{B_2})\Gamma_{ij}]f'^{n+1}_j+[\tilde{B_2}\delta_{ij}+(\frac{1}{2}-\tilde{B_2})\Gamma_{ij}]\Delta t \bm \xi\cdot\nabla f'^{n+1}_i=& \\
 +{[1-\omega(1-\bar{B_2})]\delta_{ij}+\omega(\frac{3}{2}-2\bar{B_2})\Gamma_{ij}}f'^{n}_j&\\
 -[(1-\tilde{B_2})\delta_{ij}-(\frac{1}{2}-\tilde{B_2})\Gamma_{ij}]\Delta t \bm \xi\cdot\nabla f'^{n}_i,&
\label{eq:4.5}
\end{split}
\end{equation}
where $f_j'^{n}=f'(\bm{x},\bm{\xi}_j,t_n)$ and $\Gamma_{ij}=\partial
f_i^{eq}(\bm{x},t)/\partial f_j(\bm{x},t)$. With the Fourier transform, one
can also get
\begin{equation}
 F_j(\bm k,t+\Delta t)=G_{ij}F_j(\bm k,t),
\label{eq:4.6}
\end{equation}
where  $F_j(\bm k,t)=\int f'_j(\bm x,t) \exp (-i\bm k\cdot \bm x)d\bm x$ and
the wave number $\bm k=(k_x,k_y)$. The growth matrix $\bm{G}$ can be
expressed as
\begin{equation}
\begin{split}
\bm G & =\left\{(1+\omega\bar{B_2})\bm I+\omega(\frac{1}{2}-2\bar{B_2})\bm \Gamma+[\tilde{B_2}\bm I+(\frac{1}{2}-\tilde{B_2})\bm \Gamma]r\bm S\right\}^{-1}\\
      &    \times \left\{[1-\omega(1-\bar{B_2})]\bm I+\omega(\frac{3}{2}-2\bar{B_2})\bm \Gamma-[(1-\tilde{B_2})\bm I-(\frac{1}{2}-\tilde{B_2})\bm \Gamma]r\bm S\right\},
\label{eq:4.7}
\end{split}
\end{equation}
where $r=\Delta t/\Delta x$ and $\bm S=diag(s_0,s_1,...s_q)$ depends on
$L(f)$. In the mixed difference scheme,
\begin{equation}
\begin{split}
 s_j=&l(1-\eta)(\sin\vartheta_{jx}+\sin\vartheta_{jy})+\frac{\eta}{2}[6-4\exp(-\vartheta_{jx})-4\exp(-\vartheta_{jy})\\
 & +\exp(-2l\vartheta_{jx})+\exp(-2l\vartheta_{jy})],
\label{eq:4.8}
\end{split}
\end{equation}
where $l^2=-1$, $\vartheta_{jx}=\kappa_x\xi_{jx}\Delta \chi$ and
$\vartheta_{jy}=\kappa_y\xi_{jy}\Delta \chi$.
\begin{figure}[h]
\subfigure[]{ \label{fig:stable:subfig:a}
\includegraphics[scale=0.25]{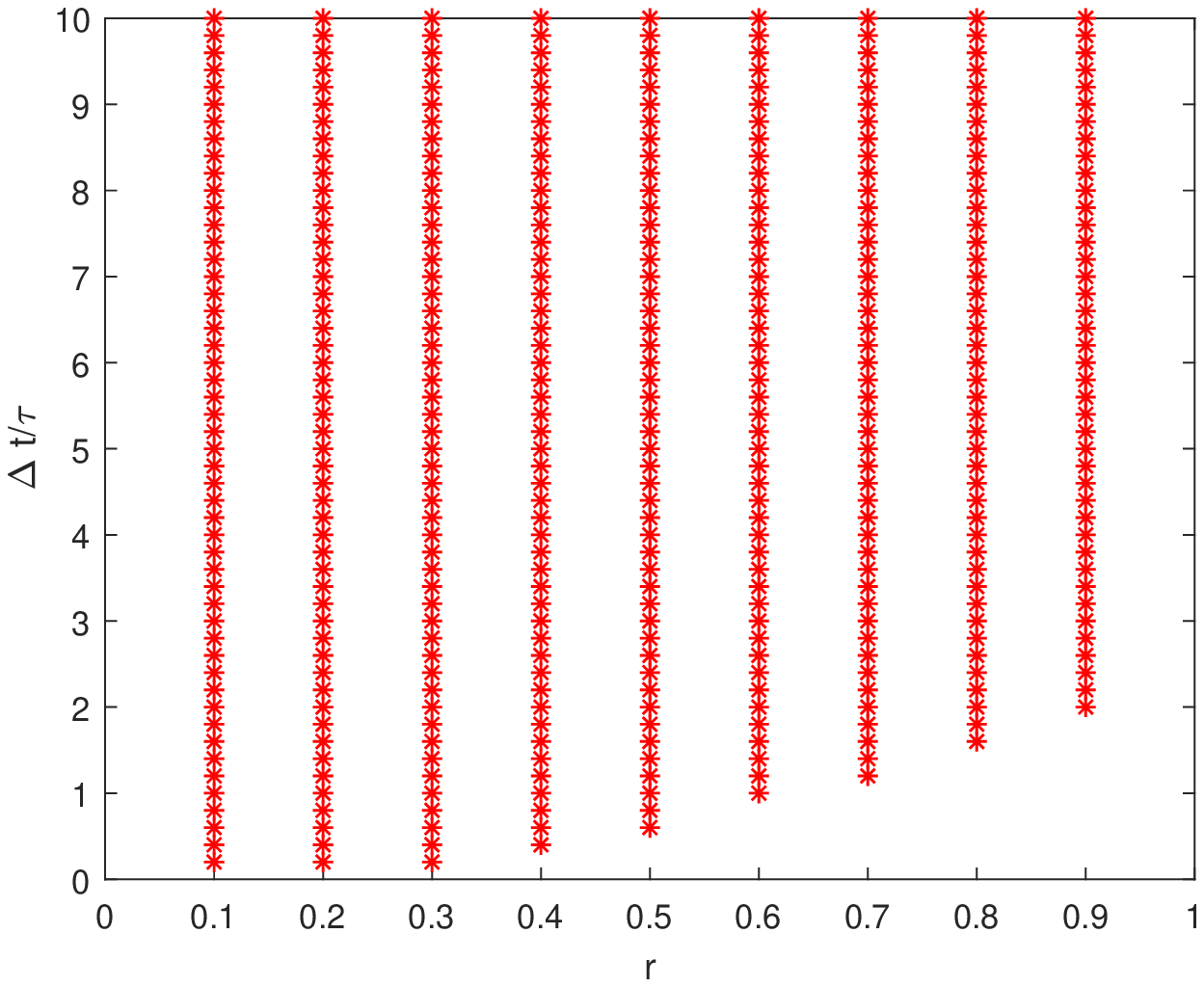}
\includegraphics[scale=0.25]{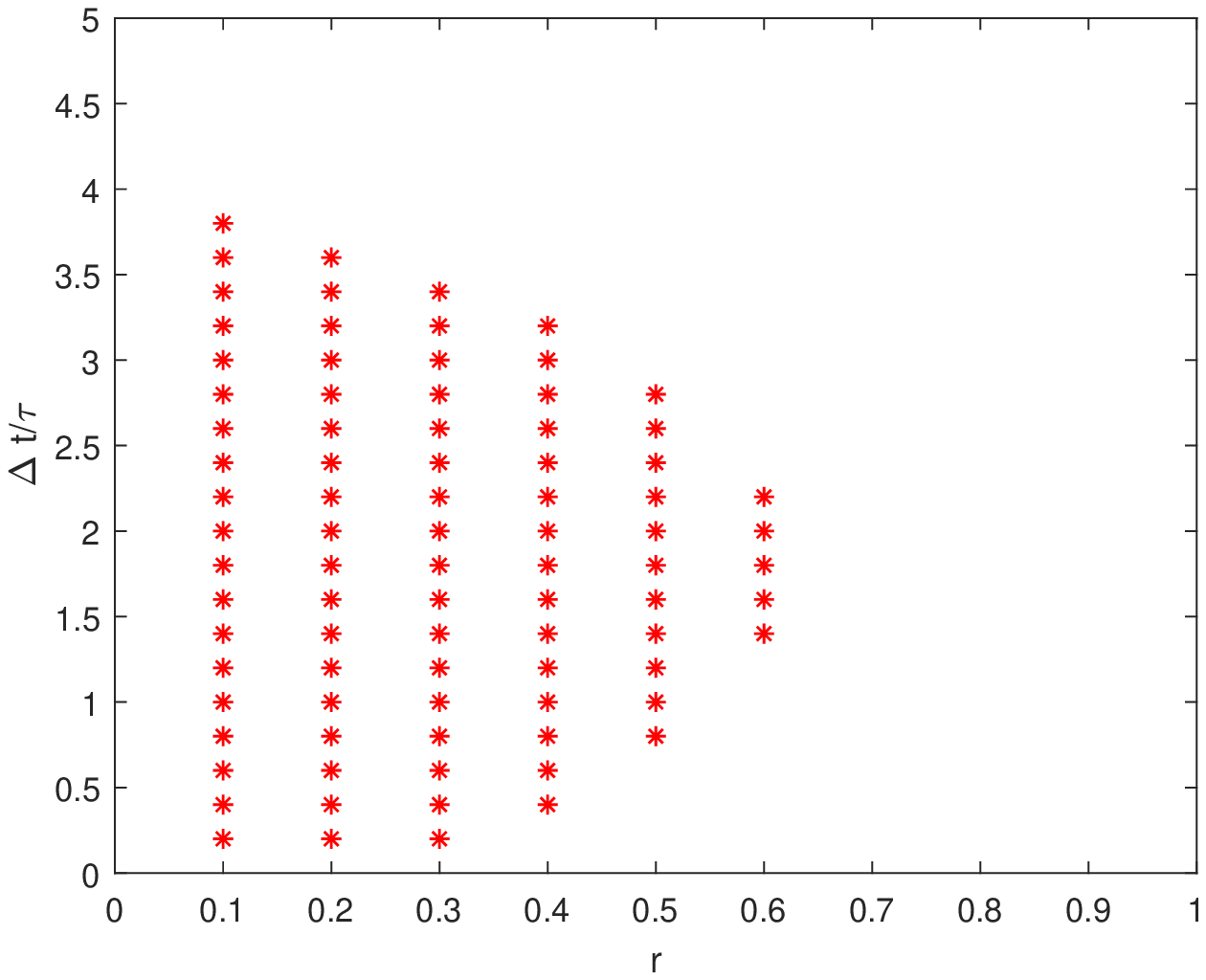}
\includegraphics[scale=0.25]{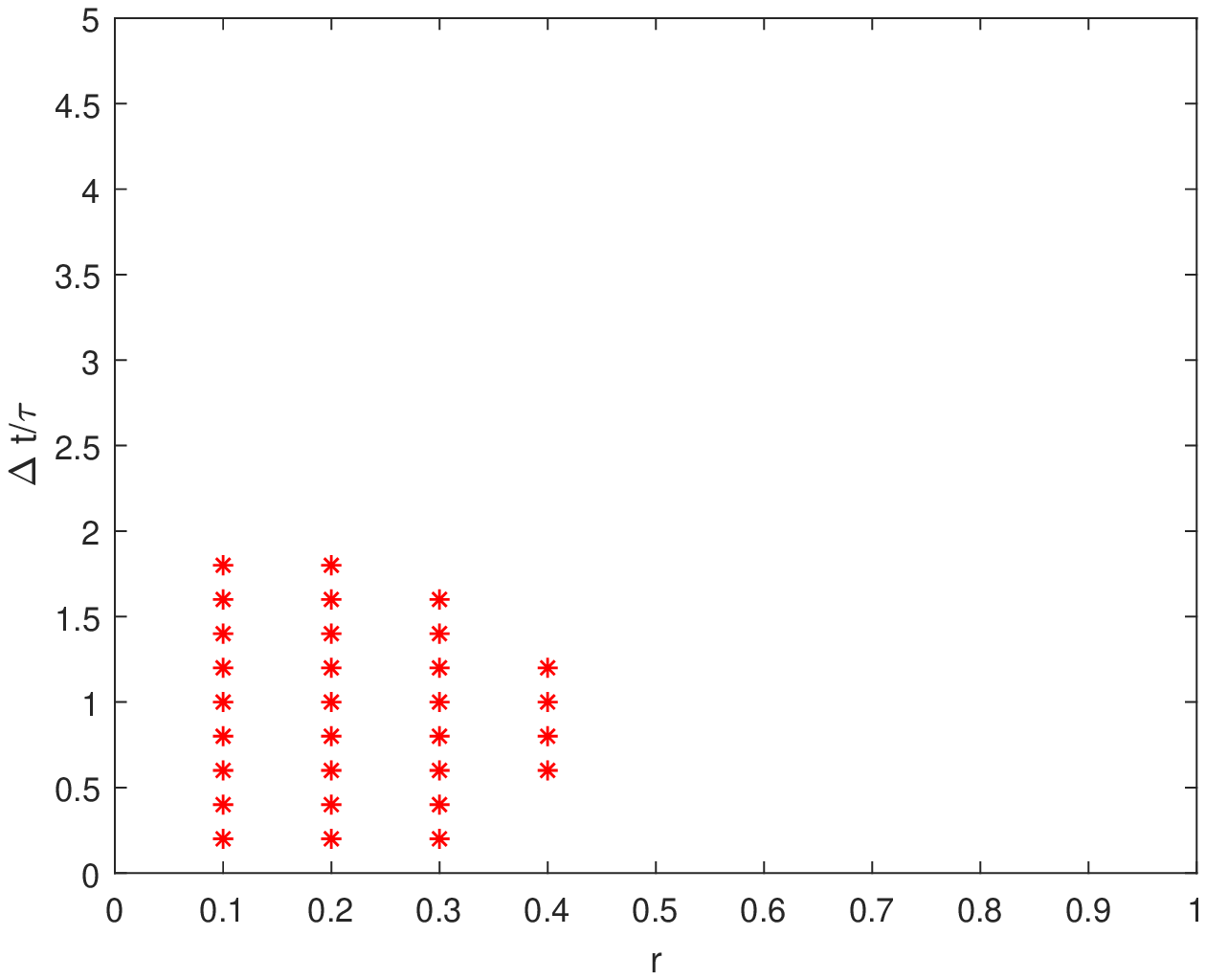}}
\subfigure[]{ \label{fig:stable:subfig:b}
\includegraphics[scale=0.25]{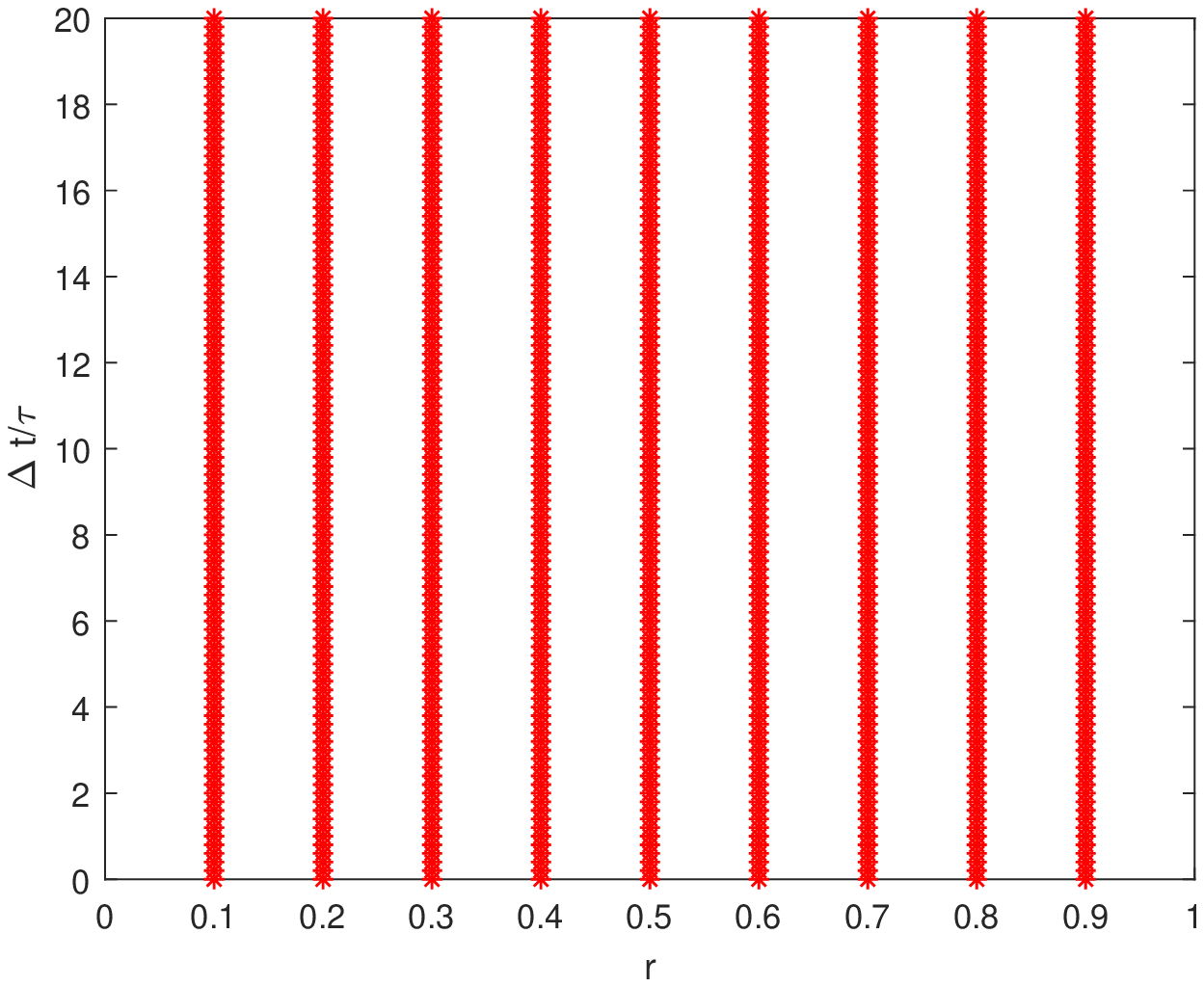}
\includegraphics[scale=0.25]{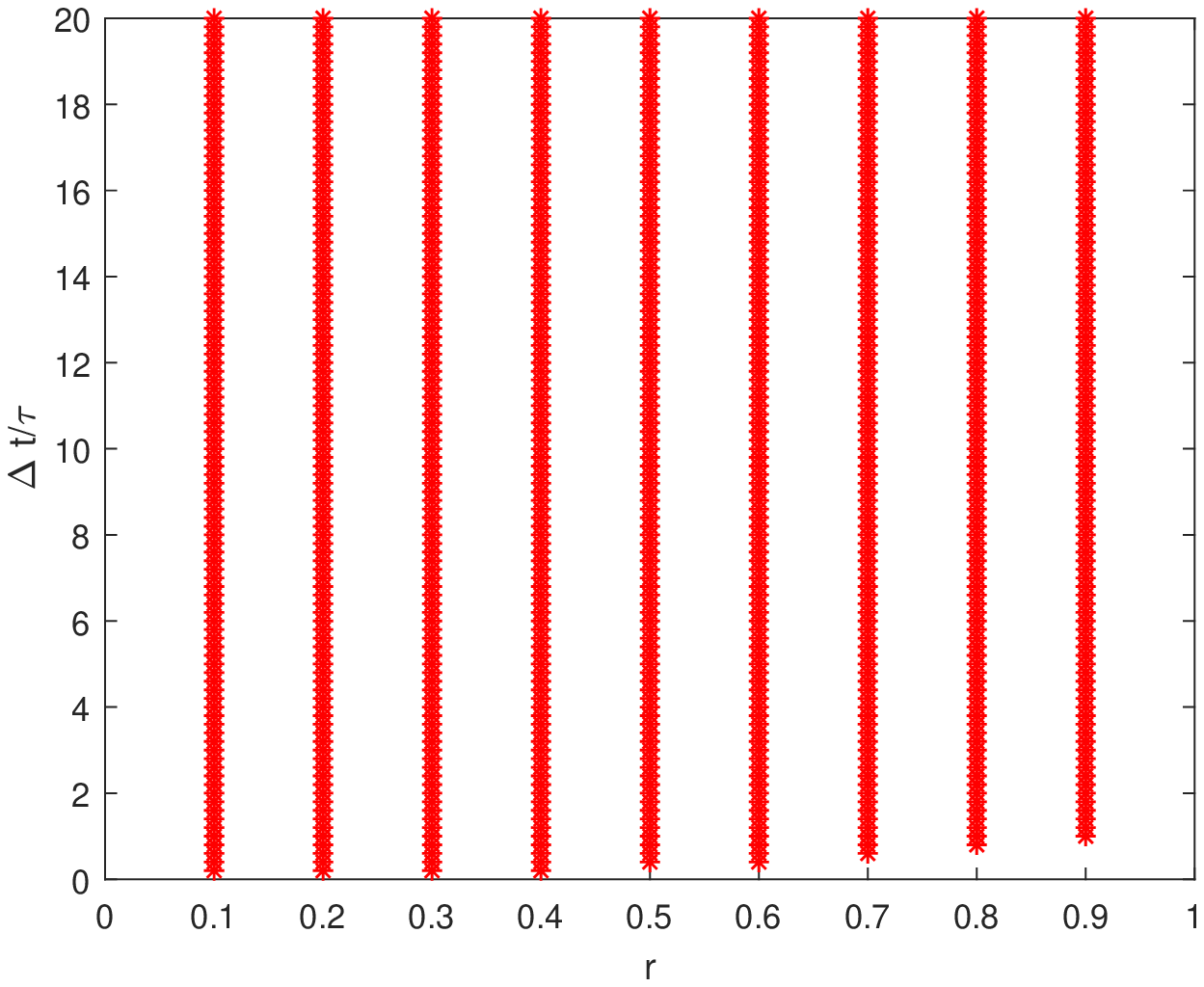}
\includegraphics[scale=0.25]{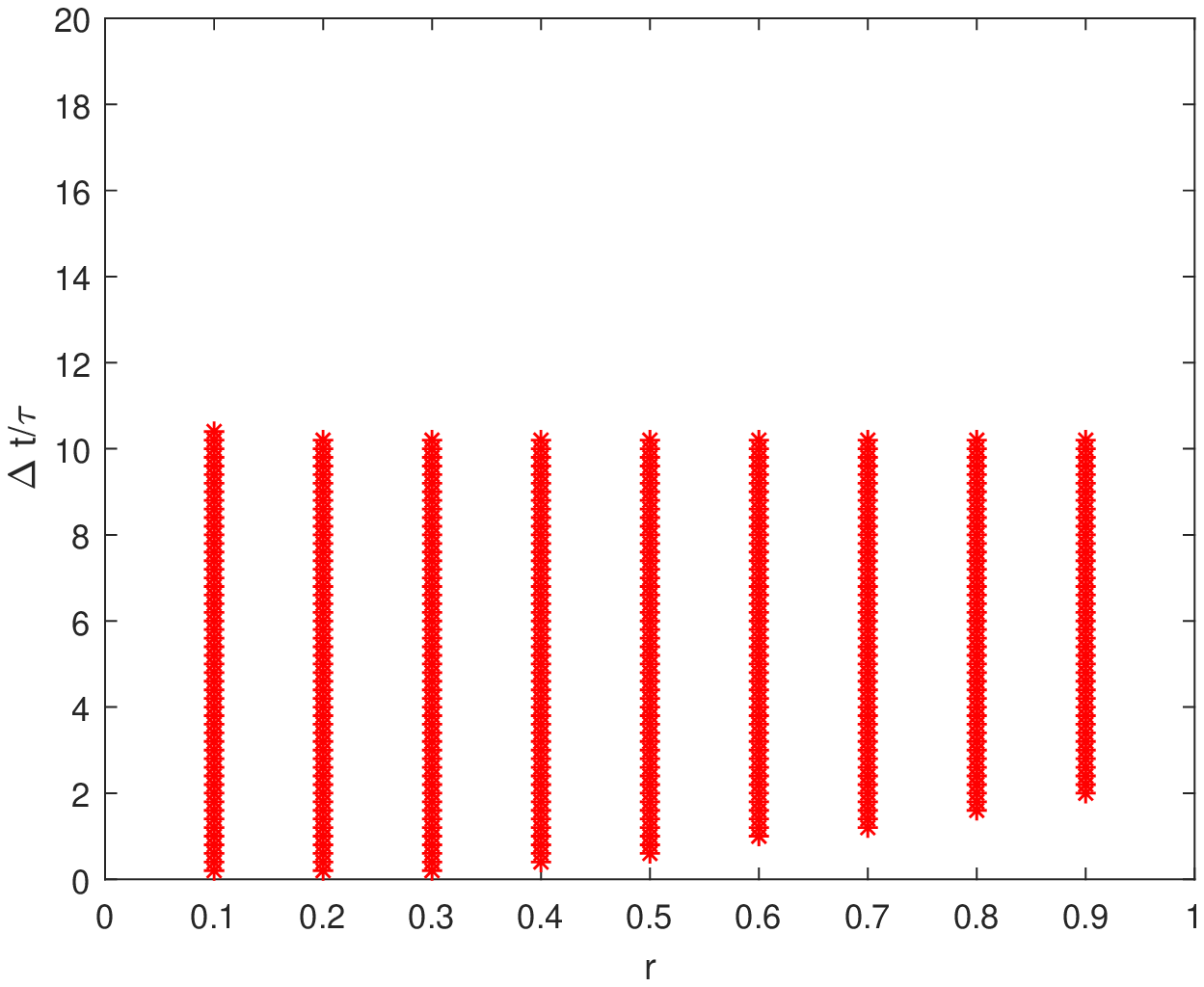}}
\subfigure[]{ \label{fig:stable:subfig:c}
\includegraphics[scale=0.25]{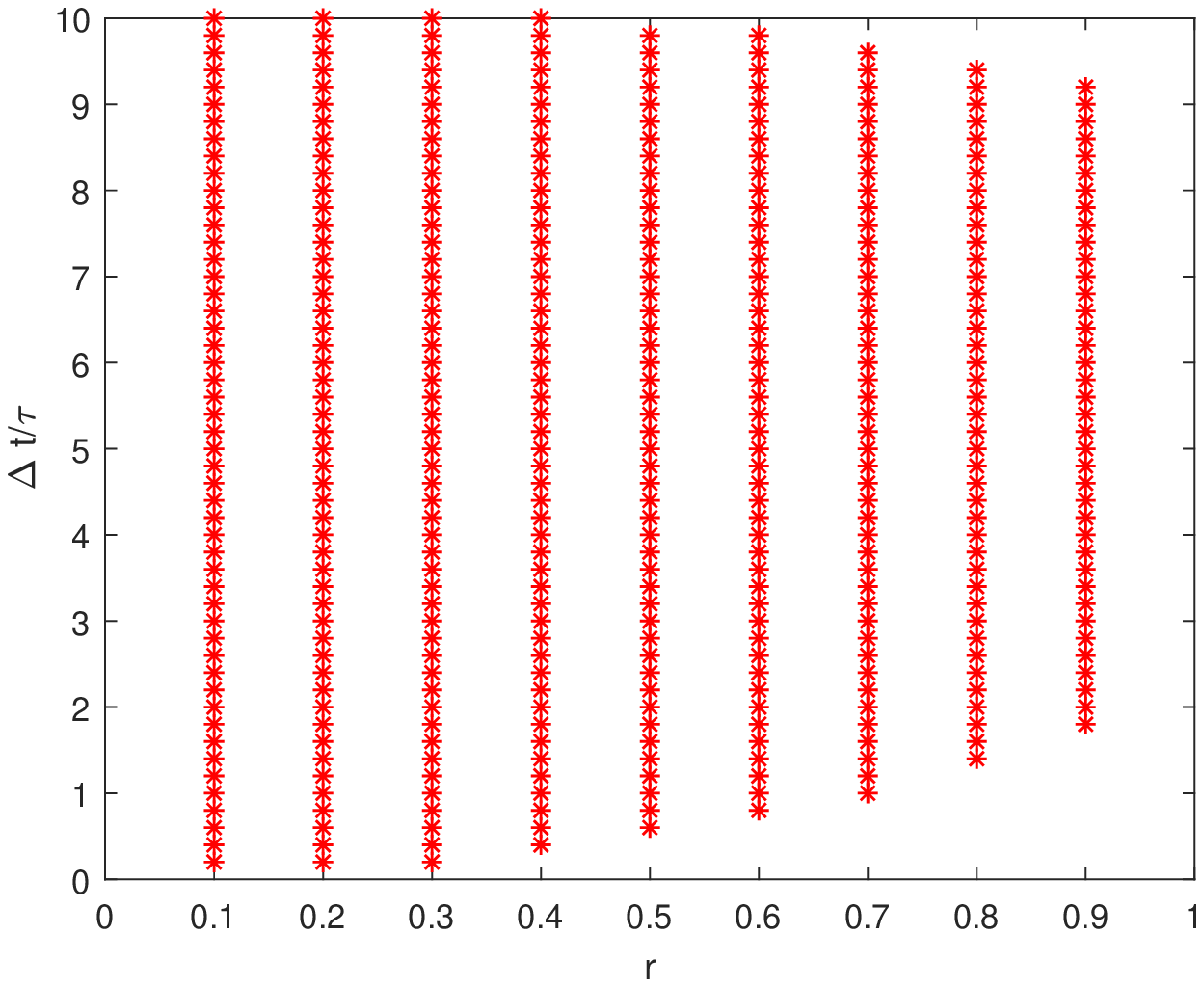}
\includegraphics[scale=0.25]{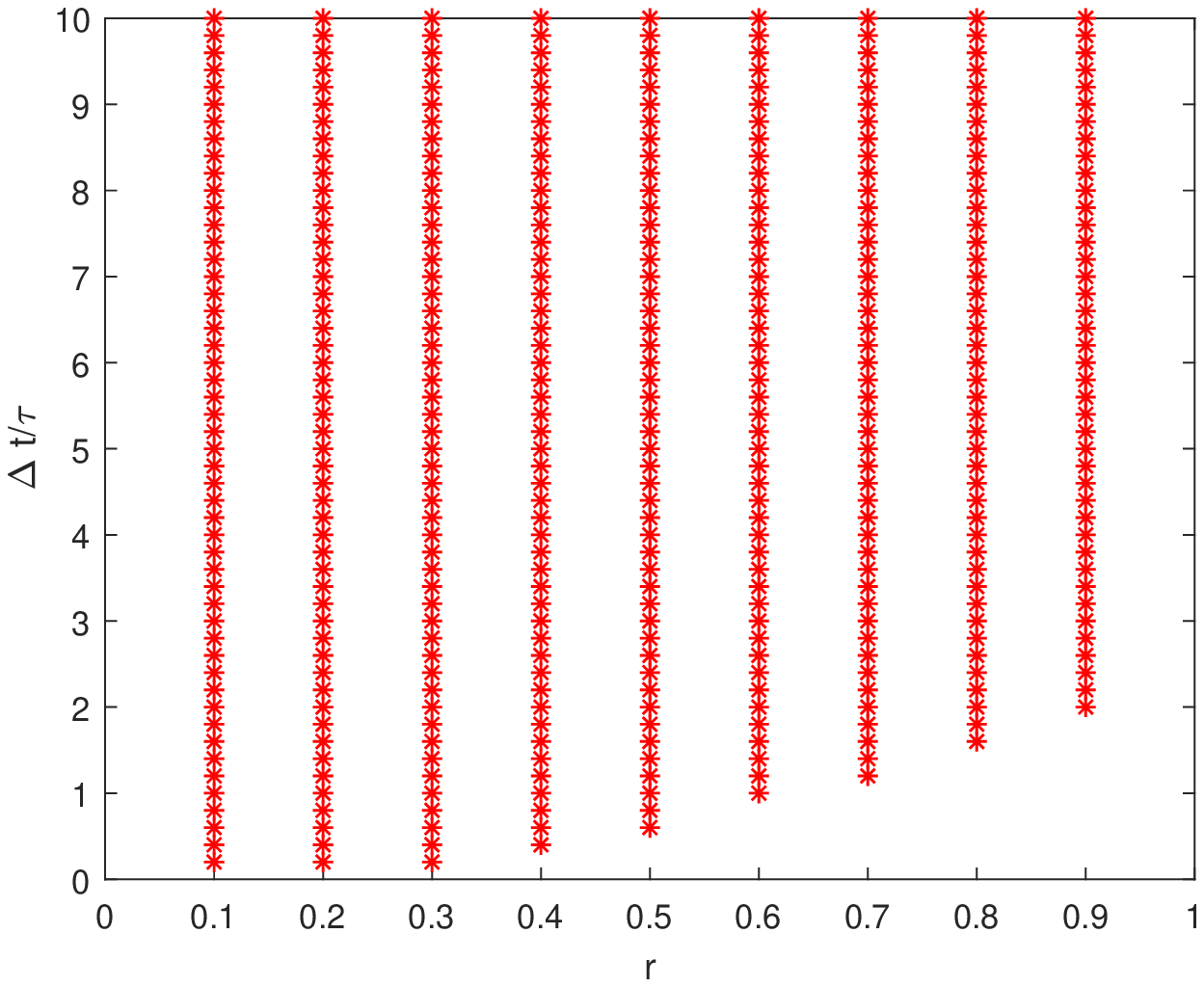}
\includegraphics[scale=0.25]{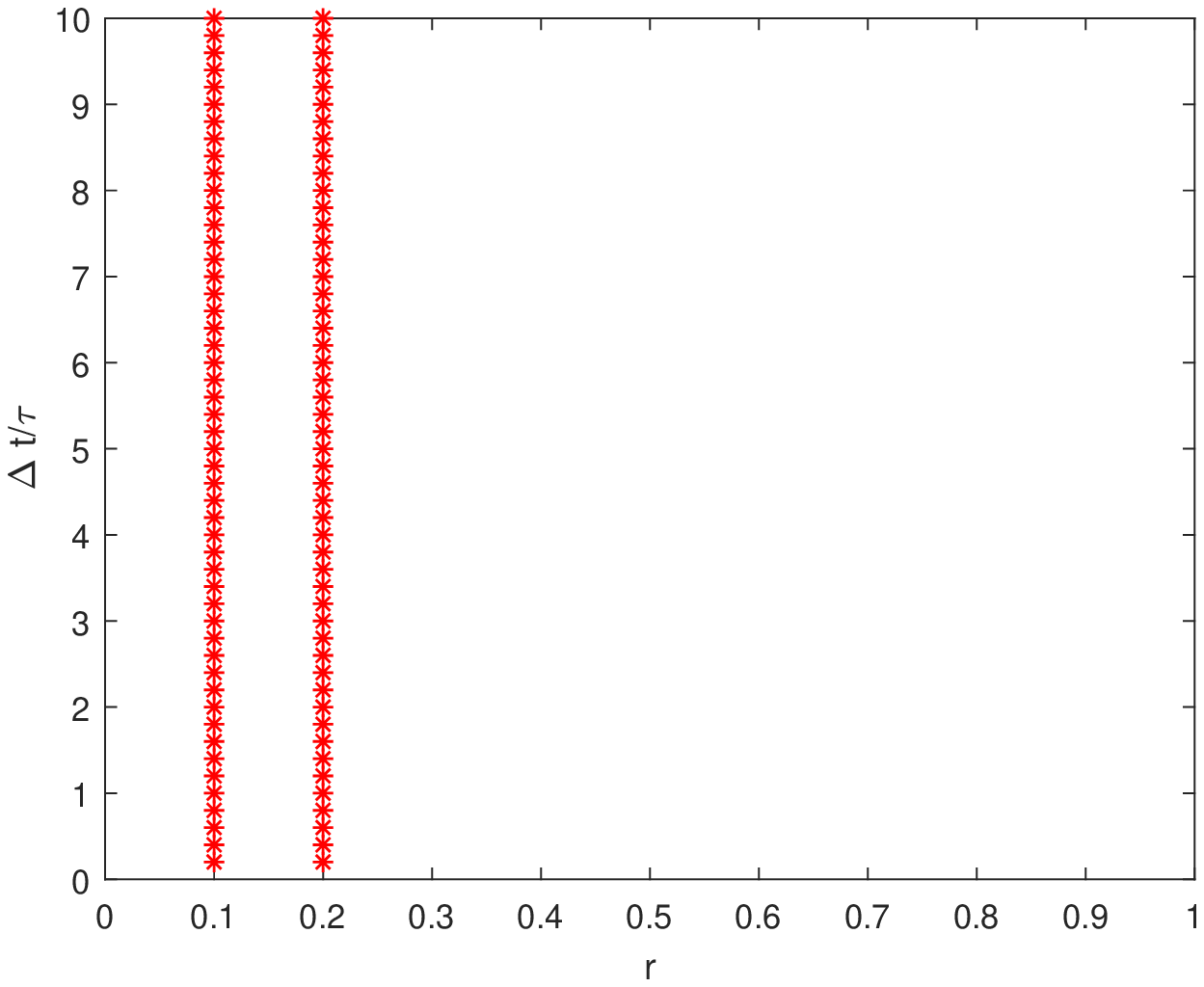}}
\caption{Stability regions of T2S2-FDLBM at different values of $\bar{B_2}$, $\tilde{B_2}$ and $\eta$. *, stable point:
 (a)$\tilde{B_2}=0$ $\eta=0.1$ and $\bar{B_2}=0.5,0.25,0$ (from left to right). (b) $\bar{B_2}=0.5$, $\eta=0.1$ and $\tilde{B_2}=0.5, 0.25, 0$ (from left to right).
 (c) $\bar{B_2}=0.5$, $\tilde{B_2}=0$ and $\eta=0, 0.1, 1$ (from left to right).}
\label{fig:stable}
\end{figure}

According to the stability condition, the spectral radius of the growth
matrix $\bm{G}$ is required to be less than 1. From Eq.~\eqref{eq:4.7}, it
is clear that the spectral radius of the matrix $\bm{G}$ depends on the
parameter of collision term ($\bar{B_2}$), the parameter of spatial gradient
term ($\tilde{B_2}$), the weight coefficient of mixed scheme ($\eta$), and
other four parameters $\bm k$, $\omega$, $r$ and $\bm u$. To perform an
analysis of the numerical stability, the parameters $\bar{B_2}$,
$\tilde{B_2}$ and $\eta$ are specified and $\bm {u}=(0.2,0.2)$, $0\leq
\kappa_\alpha\Delta \chi\leq\pi$ $(\alpha=x,y)$. As shown in
Fig.~\ref{fig:stable}, the stability region is related to $r$ and $\omega$,
and it is obvious that the present method can obtain a largest stability
region when $\bar{B_2}=0.5$, $\tilde{B_2}=0.5$ and $\eta=0.1$. Moreover,
taking account of computational efficiency, the method with $\bar{B_2}=0.5$,
$\tilde{B_2}=0$ and $\eta=0.1$ is also worthing conducting a further study.

\textbf{D. Computational sequence of two T2S2-FDLBMs}

\begin{figure}[h]
 \centering
\includegraphics[width=10cm]{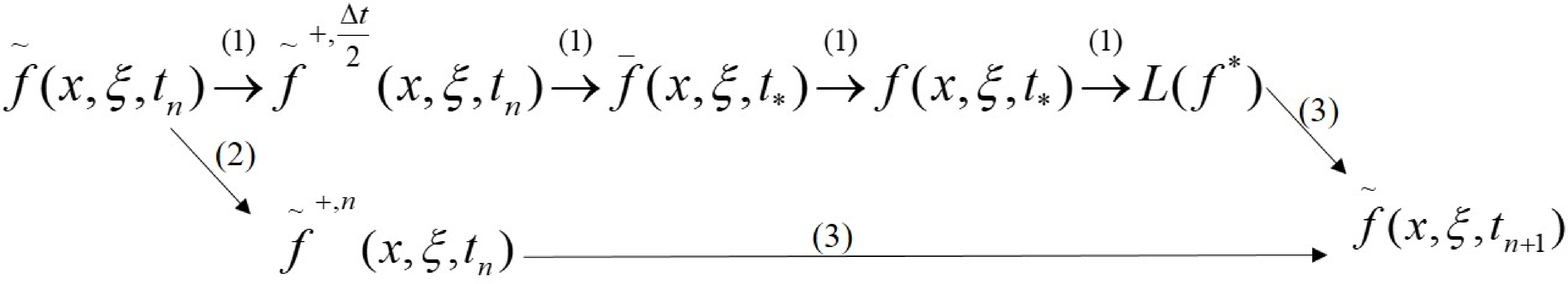}
\caption{Flow chart of T2S2-FDLBM1.} \label{fig:compute}
\end{figure}

Fig.~\ref{fig:compute} is a flow chart of T2S2-FDLBM1, in which several
steps are included.

Step (1). Estimate $L(f_i^*)$ from $\tilde{f}_i^n$ through the method of
integrating along the characteristic line,
\begin{displaymath}
 \tilde{f}(\bm{x},\bm{\xi}_i,t_n)\stackrel{\eqref{eq:2.51}}{\longrightarrow}\bar{f}^{+,h}(\bm{x},\bm{\xi}_i,t_n)\stackrel{ \eqref{eq:2.33}, \eqref{eq:2.35}}{\longrightarrow}\bar{f}(\bm{x},\bm{\xi}_i,t^*)\stackrel{\eqref{eq:2.52}}{\longrightarrow}
 f(\bm{x},\bm{\xi}_i,t^*)\stackrel{ \eqref{eq:2.27}, \eqref{eq:2.36}}{\longrightarrow}L(f_i^*),
\end{displaymath}
\begin{equation}
 \bar{f}_i^{+,h}=\frac{4\tau-\Delta t}{4\tau+2\Delta t}\tilde{f}_i+\frac{3\Delta t}{4\tau+2\Delta t}f_i^{eq},
\label{eq:2.51}
\end{equation}
\begin{equation}
 f_i=\frac{2\tau}{2\tau+h}\bar{f}_i+\frac{h}{2\tau+h}f_i^{eq}.
\label{eq:2.52}
\end{equation}

Step (2). Calculate $\tilde{f}_i^{+,n}$ from $\tilde{f}_i^n$ by
Eq.~\eqref{eq:2.20},
\begin{displaymath}
 \tilde{f}(\bm{x},\bm{\xi}_i,t_n)\stackrel{\eqref{eq:2.20}}{\longrightarrow}\tilde{f}^+(\bm{x},\bm{\xi}_i,t_n).
\end{displaymath}

Step (3). Calculate $\tilde{f}^{n+1}$ from $L(f^*)$ and $\tilde{f}^{+,n}$ by
Eq.~\eqref{eq:2.23},
\begin{displaymath}
 L(f_i^*),\tilde{f}^+(\bm{x},\bm{\xi}_i,t_n)\stackrel{\eqref{eq:2.23}}{\longrightarrow}\tilde{f}(\bm{x},\bm{\xi}_i,t_{n+1}).
\end{displaymath}

\begin{figure}[h]
 \centering
\includegraphics[width=10cm]{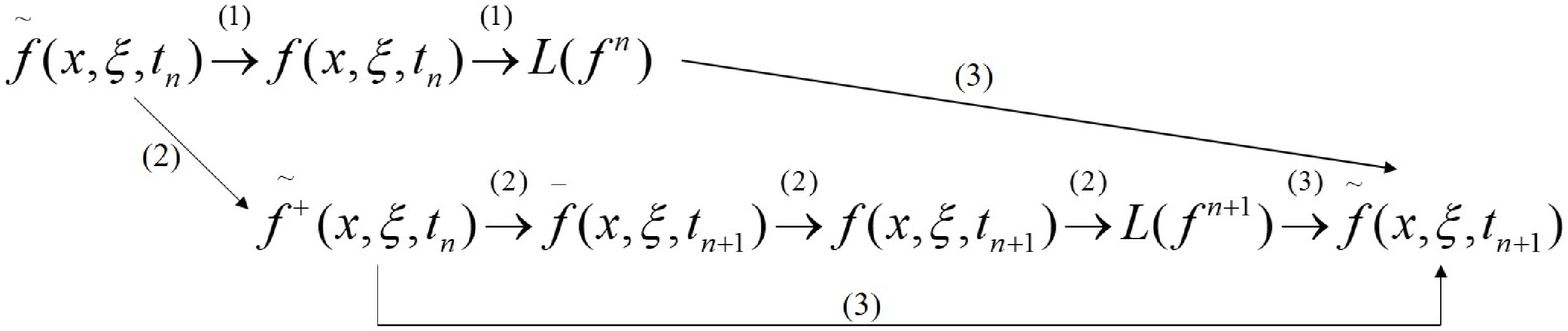}
\caption{Flow chart of T2S2-FDLBM2.} \label{fig:compute2}
\end{figure}
Similarly, we also presented computational process of T2S2-FDLBM2 in
Fig.~\ref{fig:compute2}, where the details are displayed as follows.

Step~(1). Estimate $L(f_i^n)$ from $\tilde{f}_i^n$,
\begin{displaymath}
 \tilde{f}(\bm{x},\bm{\xi}_i,t_n)\stackrel{\eqref{eq:2.52}}{\longrightarrow}f(\bm{x},\bm{\xi}_i,t_n)\stackrel{\eqref{eq:2.27},\eqref{eq:2.36}}{\longrightarrow}L(f_i^n).
\end{displaymath}

Step (2). Calculate $L(f_i^{n+1})$ from $\tilde{f}_i^n$,
\begin{displaymath}
\begin{split}
 &\tilde{f}\bm{x},\bm{\xi}_i,t_n)\stackrel{\eqref{eq:2.20}}{\longrightarrow}\tilde{f}^+(\bm{x},\bm{\xi}_i,t_n)\stackrel{\eqref{eq:2.33},
 \eqref{eq:2.35}}{\longrightarrow}\bar{f}(\bm{x},\bm{\xi}_i,t^{n+1})\\
 &\stackrel{\eqref{eq:2.52}}{\longrightarrow}f(\bm{x},\bm{\xi}_i,t_{n+1})\stackrel{\eqref{eq:2.27}, \eqref{eq:2.36}}{\longrightarrow}L(f_i^{n+1}).
\end{split}
\end{displaymath}

Step (3). Calculate $\tilde{f}_i^{n+1}$ from $L(f_i^n)$, $L(f_i^{n+1})$ and
$\tilde{f}_i^{+,n}$ by Eq.~\eqref{eq:5.23},
\begin{displaymath}
 L(f_i^n),L(f_i^{n+1}),\tilde{f}^+(\bm{x},\bm{\xi}_i,t_n)\stackrel{\eqref{eq:5.23}}{\longrightarrow}\tilde{f}(\bm{x},\bm{\xi}_i,t_{n+1}).
\end{displaymath}

In the implementation, $\bar{f}_i=\tilde{f}_i$ and
$\bar{f}_i^{+,h}=\tilde{f}_i^+$ when $h=\Delta t$. Here it should be noted
that the computational cost of T2S2-FDLBM2 is larger than T2S2-FDLBM1. The
main reason is that two terms $L(f_i^n)$ and $L(f_i^{n+1})$ should be
calculated in T2S2-FDLBM2, while only $L(f_i^n)$ needs to be calculated in
T2S2-FDLBM1.

\section{\label{sec:level3}Numerical simulation}
In this section, the Taylor vortex flow, the Poiseuille flow and the
lid-driven flow will be used to test the two T2S2-FDLBMs.

Unless otherwise stated, in our simulations the equilibrium distribution
function is adopted to initialize the distribution function, and the time
step $\triangle t$ is given by the $CFL$ condition number,
\begin{equation}
 \triangle t=CFL\frac{\triangle x}{\xi},
\label{eq:3.1}
\end{equation}
where $\triangle x$ is the minimum grid scale, and $\xi=\max|\bm{\xi}_i|$.
$CFL$ condition number is an important parameter to evaluate the stability
and convergence of method. The collision term $\Omega$ can be approximated
by the simple single-relaxation-time Bhatnagar-Gross-Krook model,
\begin{equation}
\ \Omega(f_i)=-\frac{{1}}{\tau}[f_i-f_i^{eq}],
\label{eq:2.4}
\end{equation}
where the equilibrium distribution function $f_i^{eq}$ in SLBM is defined as
\begin{equation}
\ f_i^{eq}=\omega_i\rho\left[1+\frac{{\bm{\xi}_i\cdot \bm{u}}}{c_s^2}+
\frac{{\bm{uu}:(\bm{\xi}_i \bm{\xi}_i -c_s^2\textbf{I})}}{2c_s^4}\right],
\label{eq:3.2}
\end{equation}
and $f_i^{eq}$ in He-Luo model \cite{he1997lattice}can be also combined with
this two T2S2-FDLBMs. The discrete particle velocities and corresponding
weights are dependent on the lattice structure. For example,\\
D1Q3:
\begin{equation}
 \begin{split}
&\bm{\xi}_i=(0,1,-1)\xi,\\
&\omega_0=\frac{2}{3}, \omega_1=\omega_2=\frac{1}{6}, c_s^2=\frac{\xi^2}{3},
  \end{split}
\end{equation}
D2Q9:
\begin{equation}       
\begin{split}
&\bm{\xi}_i=\left(                 
  \begin{array}{ccccccccc}   
    0 & 1 & 0 & -1 &  0 & 1 & -1 & -1 &  1\\  
    0 & 0 & 1 &  0 & -1 & 1 &  1 & -1 & -1\\  
  \end{array}
\right)\xi, \\               
&\omega_i=\left(                 
  \begin{array}{ccccccccc}   
    \dfrac{4}{9} & \dfrac{1}{9} & \dfrac{1}{9} & \dfrac{1}{9} &  \dfrac{1}{9} & \dfrac{1}{36} & \dfrac{1}{36} & \dfrac{1}{36} &  \dfrac{1}{36}\\  
  \end{array}
\right), c_s^2=\frac{\xi^2}{3}.
\end{split}
\end{equation}
The macroscopic density $\rho$ and velocity $\textbf{u}$ can be obtained
from the distribution function,
\begin{equation}
\ \rho=\Sigma \tilde{f_i},\quad   \rho\textbf{u}=\Sigma \bm{\xi}_i \tilde{f_i}+\frac{1}{2}\Delta t\rho \bar{F},
\label{eq:3.3}
\end{equation}
where $\bar{F}$ represents the external force. According to the previous
work \cite{guo2002discrete}, the force term in T2S2-FDLBM can be expressed
as
\begin{equation}
\ F_i=\bar{F}\cdot (\xi_i-u)f_i^{eq}/(RT).
\label{eq:3.12}
\end{equation}

\textbf{A. The Taylor vortex flow}

The two-dimensional Taylor vortex flow is a periodic problem, and it is
widely used to test the accuracy of the model. The analytical solution of
the Taylor vortex flow  is given by
\begin{equation}
\begin{split}
u & =  -u_0\cos(k_1x)\sin(k_2y)\exp\left[ -\nu(k_1^2+k_2^2)t\right],\\
v & =  u_0\frac{k_1}{k_2}\sin(k_1x)\cos(k_2y)\exp\left[ -\nu(k_1^2+k_2^2)t\right],\\
p & = p_0-\frac{u_0^2}{4}\left[ \cos(2k_1x)+\frac{k_1^2}{k_2^2}\cos(2k_2y)\right] \exp\left[ -\nu(k_1^2+k_2^2)t\right],
\end{split}
\label{eq:3.4}
\end{equation}
where $u$ and $v$ are horizontal and vertical velocities of the fluid, $p$
is the pressure. The computational domain of the problem is set as $-\pi\leq
x,y\leq\pi$, the mesh size is chosen to be $Nx\times Ny=32\times 128$,
$k_1=1.0$, $k_2=4.0$, $u_0$ is set to be $0.01$. The time step is chosen to
be $\pi/640$, and the shear viscosity $\nu$ is set as 0.001. The density can
be initialized by $\rho=\rho_0+\delta p/c_s^2$, the average density
$\rho_0=p_0/c_s^2$ and $\delta p=p-p_0$. For this time-dependence problem,
the initial distribution function is given by \cite{guo2003explicit},
\begin{equation}
 \tilde{f}_i(\bm{x},\bm{\xi},t_0)=f_i^{eq}(\bm{x},\bm{\xi},t_0)-\frac{\rho_0 \omega_i \tau (2+\Delta t)}{2c_s^2}\bm{\xi}_i\bm{\xi}_i:\nabla \bm{u}(\bm{x},\xi,t_0),
\label{eq:3.5}
\end{equation}
where $\rho_0$ and $u$ are determined by analytical solution.

\begin{figure}[h]
\subfigure[]{ \label{fig:Taylor:a}
\includegraphics[width=2.3in]{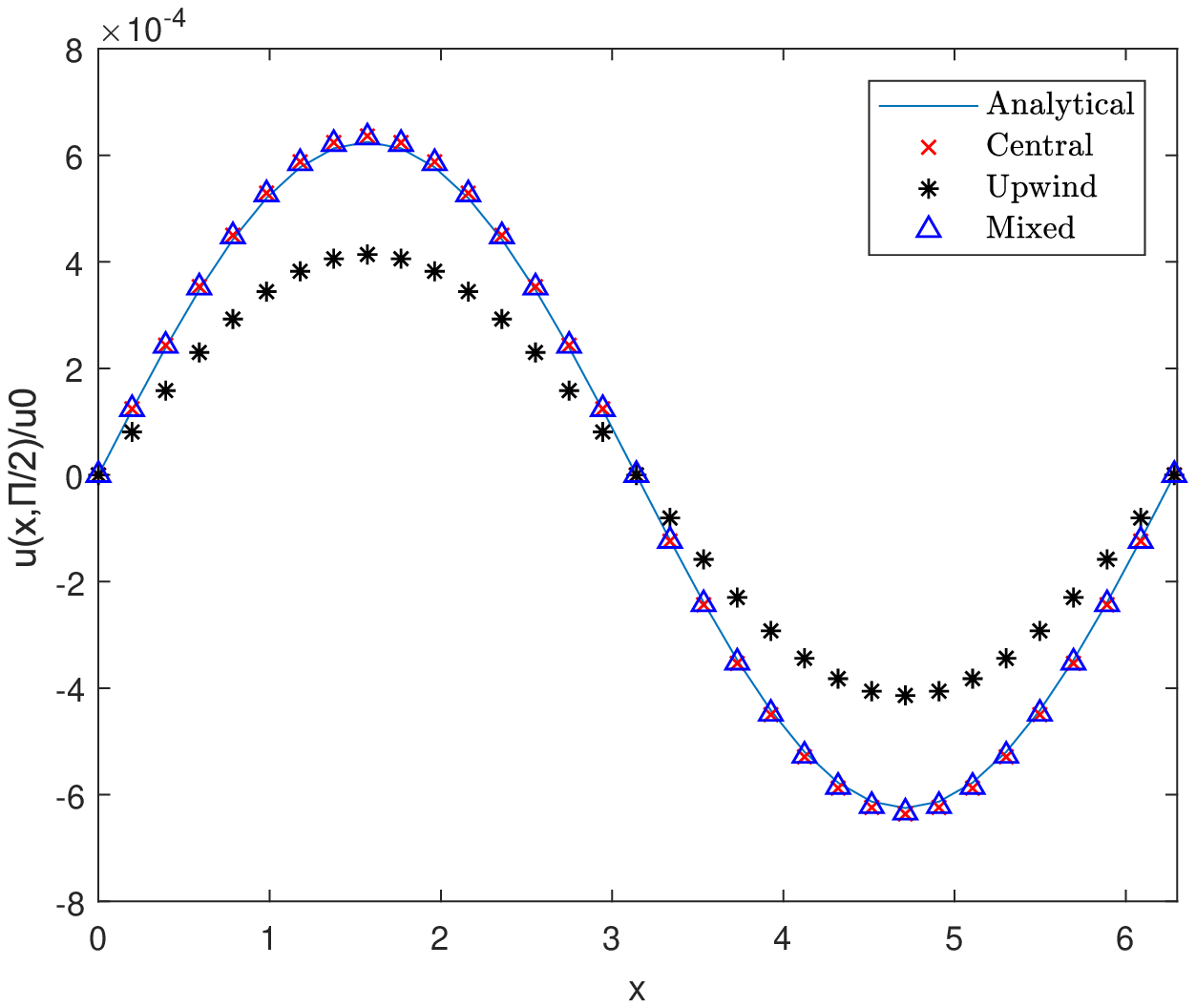}}
\subfigure[]{ \label{fig:Taylor:b}
\includegraphics[width=2.3in]{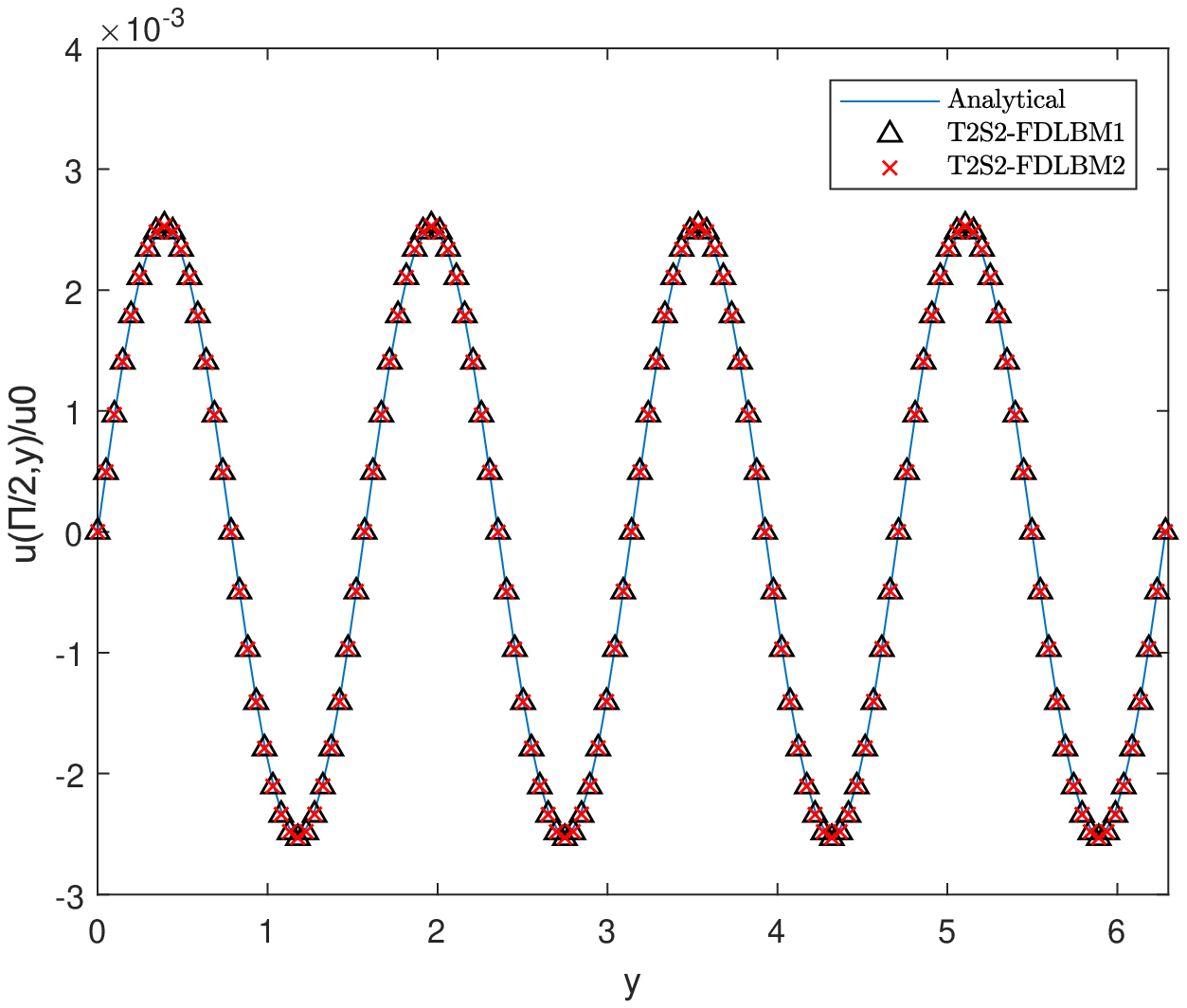}}
\caption{Velocity profiles at $t_c=ln[2\nu(k_1^2+k_2^2)]$
[(a) $u$ along the vertical centerline, (b) $v$ along the horizontal centerline].}
\label{fig:Taylor}
\end{figure}

Two T2S2-FDLBMs are first used to simulate the Taylor vortex flow, and the
gradient term is discretized by three difference schemes: second-order
upwind, central and mixed difference schemes ($\eta=0.01$). The results of
three difference schemes are shown in Fig.~\ref{fig:Taylor}. It can be
observed from Fig.~\ref{fig:Taylor:a} that the T2S2-FDLBM1 with the up-wind
difference scheme has a significant error, while the T2S2-FDLBM1 with the
central difference or mixed difference scheme agree well with the analytical
solution. Besides, from Fig.~\ref{fig:Taylor:b}, one can also find that the
results of T2S2-FDLBM1 and T2S2-FDLBM2 are in good agreement with the
analytical solution when $CFL=0.1$. However, Table~\ref{table1} shows that
difference schemes play an important role in the T2S2-FDLBM1. The up-wind
difference scheme has a serious error, and the error of mix difference
scheme is smaller than that of central difference scheme. This phenomenon is
consistent with that of T1S2-FDLBM.

\begin{table}[!htbp]
\caption{Errors of FDLBM with three difference schemes.}
\label{table1}
  \centering
  \vspace{1ex}
\resizebox{\textwidth}{!}{
\begin{tabular}{clccccccc}
\hline\hline
\            & $\qquad$  & T1S2-FDLBM  & $\qquad$ &   T2S2-FDLBM1    & $\qquad$ &  T1S2-FDLBM   & $\qquad$ &  T2S2-FDLBM1         \\
\            & $\qquad$  & $E(u)$        & $\qquad$ &    $E(u)$         & $\qquad$ &  $E(v) $       & $\qquad$ & $ E(v)   $            \\
\hline
\  up-wind & $\qquad$  & $0.186$       & $\qquad$ &   $0.185$         & $\qquad$ &  $0.184$        & $\qquad$ &  $0.185$              \\
\  central & $\qquad$  & $0.00894$     & $\qquad$ &   $0.00883$       & $\qquad$ &  $0.00884$      & $\qquad$ &  $0.00878$            \\
\  mixed  & $\qquad$  & $0.00678$     & $\qquad$ &   $0.00667$       & $\qquad$ &  $0.00666$      & $\qquad$ &  $0.00660$            \\
\hline\hline
\end{tabular}
}
\end{table}

In order to test the accuracies of T2S2-FDLBM1 and T2S2-FDLBM2, different
grid sizes ($N_x\times N_y=16\times 64,32\times 128,48\times 192,64\times
256, 80\times 320,96\times 384$) and the global relative error ($GRE$) of
velocity at $t=t_c$ are considered,
\begin{equation}
E(u) = \frac{\sqrt{\sum_{i,j} |u_{i,j}-u_{i,j}'|^2}}{\sqrt{\sum_{i,j} |u_{i,j}'|^2}},
\label{eq:3.6}
\end{equation}
where $u_{ij}$ and $u_{ij}'$ are numerical and analytical solutions.

As seen from Table~\ref{table:Taylor_GRE}, the errors obtained from
T1S2-FDLBM, T2S2-FDLBM1 and T2S2-FDLBM2 are almost the same, and all of them
have a second-order accuracy in space. This can be explained by the fact
that the three models use the same mixed difference scheme ($\eta=0.01$) to
deal with the gradient terms.
\begin{table}[!htbp]
\caption{$GRE$s and convergence order of FDLBM with $\Delta t=\pi/640$.}
  \centering
  \vspace{1ex}
\resizebox{\textwidth}{!}{
\begin{tabular}{ccccccccccccccc}
\hline\hline
\  Model    & $\quad$  & $N_x\times N_y$      & $\quad$  & $16\times 64$        & $\quad$ &  $32\times 128$      & $\quad$  &   $48\times 192$        & $\quad$  & $64\times 256$       & $\quad$  &  $80\times 320$       & $\quad$ &  $96\times 384$          \\
\hline

\  T1S2-       & $\quad$  & $E(u)$  & $\quad$  & $1.92\times 10^{-2}$ & $\quad$ & $6.65\times 10^{-3}$ & $\quad$  &   $ 3.21\times 10^{-3}$ & $\quad$  & $1.88\times 10^{-3}$ & $\quad$  &  $1.24\times 10^{-3}$ & $\quad$ & $9.07\times 10^{-4}$     \\

\  FDLBM       & $\quad$  & order & $\quad$  & $--$                 & $\quad$ & $1.5326$             & $\quad$  & $1.7981$                & $\quad$  & $ 1.8632$            & $\quad$  & $1.8437$              & $\quad$ &  $1.7349$     \\

\ T2S2-      & $\quad$  & $E(u)$  & $\quad$  & $1.93\times 10^{-2}$ & $\quad$ & $6.66\times 10^{-3}$ & $\quad$  &   $ 3.21\times 10^{-3}$ & $\quad$  & $1.88\times 10^{-3}$ & $\quad$  &  $1.24\times 10^{-3}$ & $\quad$ & $8.94\times 10^{-4}$     \\

\  FDLBM1       & $\quad$  & order & $\quad$  & $--$                 & $\quad$ & $1.5318$             & $\quad$  & $1.7983$                & $\quad$  & $ 1.8691$            & $\quad$  & $1.8645$              & $\quad$ &  $1.7830$     \\
\  T2S2-       & $\quad$  &$ E(u)$ & $\quad$  & $1.93\times 10^{-2}$ & $\quad$ & $6.78\times 10^{-3}$ & $\quad$  &   $ 3.32\times 10^{-3}$ & $\quad$  & $1.98\times 10^{-3}$ & $\quad$  &  $1.34\times 10^{-3}$ & $\quad$ & $9.83\times 10^{-4}$     \\

\  FDLBM2       & $\quad$  & order & $\quad$  & $--$                 & $\quad$ & $1.5157$             & $\quad$  & $1.7566$                & $\quad$  & $ 1.7966$            & $\quad$  & $1.7679$              & $\quad$ &  $1.6841$     \\
\hline\hline
\end{tabular}
}\label{table:Taylor_GRE}
\end{table}
\begin{figure}[h]
\centering
\includegraphics[scale=0.5]{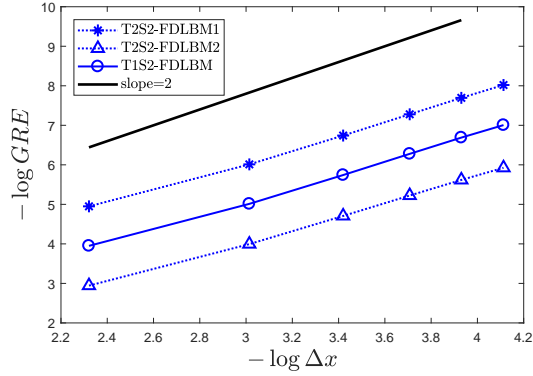}
\caption{$GRE$ of FDLBM at different gride sizes.}\label{fig:sorder}
\end{figure}

To analyze the stability of the model, we also performed some simulations
with different values of $CFL$ condition number. Table~\ref{table:CFL} shows
the $GRE$s of T1S2-FDLBM, T2S2-FDLBM1 and T2S2-FDLBM2. From this table, it
can be observed that T1S2-FDLBM is unstable when $CFL$ condition number is
more than $0.3$. Compared to T2S2-FDLBM2, T2S2-FDLBM1 works well and
maintains a small error even when $CFL=0.9$. These results indicate that the
two T2S2-FDLBMs are more stable than the T1S2-FDLBM, and simultaneously,
T2S2-FDLBM1 is more suitable to simulate Taylor vortex flow as the $CFL$
condition number increases. It should be noted from the Eq.~\eqref{eq:3.1}
that $\Delta t$ is proportional to $CFL$ condition number, it means that the
value of $CFL$ condition number is related to computational efficiency. For
this reason, we further tested the computational efficiency of four methods,
i.e., T2S2-FDLBM1, T2S2-FDLBM2, T1S2-FDLBM and SLBM, and presented the
results in Table~\ref{table:Taylor_comparison}. It is found that the CPU
time of SLBM, T1S2-FDLBM and T2S2-FDLBM2 are $64$ times, $4.6$ times and
$10$ times as long as T2S2-FDLBM1 under the similar error. In this example,
one can find that T2S2-FDLBM1 has good stability and high computational
efficiency.
\begin{table}[!htbp]
\caption{$GRE$s of FDLBM with different values of $CFL$ condition number.}
  \centering
  \vspace{1ex}
\resizebox{\textwidth}{!}{
\begin{tabular}{ccccccccccccccccccccc}
\hline\hline
\   $CFL$    & $\quad$  & $0.1$         & $\quad$ &   $0.2$          & $\quad$ &  $0.3$           & $\quad$  &  $0.4 $
             & $\quad$  &  $0.5 $       & $\quad$ &  $0.6 $          & $\quad$ &  $0.7 $          & $\quad$  &  $0.8 $
             & $\quad$  &  $0.9 $ \\
\hline
\ T1S2-FDLBM $GRE$ & $\quad$  &  $0.0063$     & $\quad$ &   $0.0127$       & $\quad$ &             & $\quad$  &  $- $
             & $\quad$  &  $- $         & $\quad$ &  $- $            & $\quad$ &  $- $            & $\quad$  &  $- $
             & $\quad$  &  $- $ \\
\ T2S2-FDLBM1 $GRE$ & $\quad$  &  $0.0064$     & $\quad$ &   $0.0128$       & $\quad$ &  $0.0191$             & $\quad$  &  $0.0255 $
             & $\quad$  &  $0.0321 $         & $\quad$ &  $0.0388 $            & $\quad$ &  $0.0460$            & $\quad$  &  $0.0537$
             & $\quad$  &  $0.0625 $ \\
\ T2S2-FDLBM2 $GRE$ & $\quad$  &  $0.0058$     & $\quad$ &   $0.0026$       & $\quad$ &  $0.0058$             & $\quad$  &  $0.0161 $
             & $\quad$  &  $0.0304 $         & $\quad$ &  $0.0490 $            & $\quad$ &  $0.0727$            & $\quad$  &  $0.1027$
             & $\quad$  &  $1.0104 $ \\
\hline\hline
\end{tabular}
}\label{table:CFL}
\end{table}
\begin{table}[!htbp]
\centering
\caption{A comparison of four models for the Taylor vortex flow at $t=2t_c$.}
\resizebox{\textwidth}{!}{
\begin{tabular}{m{7em}<{\centering}|m{1em}m{5em}<{\centering}m{5em}<{\centering}m{1em}m{5em}<{\centering}m{5em}<{\centering}m{1em}m{5em}<{\centering}m{5em}<{\centering}
m{1em}m{5em}<{\centering}m{5em}<{\centering}}
\hline\hline
\  model  &  & \multicolumn{2}{c}{T2S2-FDLBM1} &  & \multicolumn{2}{c}{T2S2-FDLBM2}  &     &    \multicolumn{2}{c}{T1S2-FDLBM}            &  &  \multicolumn{2}{c}{SLBM}               \\
\cline{3-4}\cline{6-7}\cline{9-10}\cline{12-13}
\  grid & & $32\times 128$ & $32\times 128$ & & $32\times 128$ & $32\times 128$ & & $32\times 128$ & $32\times 128$ & & $512\times 512$ & $128\times 128$   \\
\  $CFL$ & & $0.8$ & $0.9$ & & $0.1$ & $0.6$ & & $0.1$ & $0.2$ & & -- & --    \\
\  $\Delta t$ & & $0.0393$ & $0.0442$ & & $0.0049$ & $0.0294$ & & $0.0049$ & $0.0098$ & & $0.0123$ & $0.0491$  \\
\  iterative times & & $2076$ & $1845$ & & $16612$ & $2768$ & & $16612$ & $8306$ & & $6645$ & $1661$  \\
\cline{3-4}\cline{6-7}\cline{9-10}\cline{12-13}
\  CPU time & & $2.12$ & $1.91$ & & $19.04$ & $3.31$ & & $8.91$ & $4.54$ & & $122.88$ & $2.05$  \\
\  ratio & & $1.1099$ & $1.0000$ & & $9.9686$ & $1.7330$ & & $4.6649$ & $2.3770$ & & $64.3351$ & $1.0733$  \\
\cline{3-4}\cline{6-7}\cline{9-10}\cline{12-13}
\  $GRE$ $\times 10^{-2}$ & & $1.3095$ & $1.3228$ & & $1.1643$ & $9.4104$ & & $1.3412$ & $1.3226$ & & $1.2701$ & $4.8907$  \\
\  ratio & & $0.9900$ & $1.0000$ & & $0.8802$ & $7.1140$ & & $1.0139$ & $ 0.9998$ & & $0.9602$ & $3.6972$  \\
\hline\hline
\end{tabular}
}\label{table:Taylor_comparison}
\end{table}

\textbf{B. The two-dimensional Poiseuille flow}

Considering the Poiseuille flow driven by a constant external force in a
two-dimensional channel, the analytical solution of velocity can be express
as
\begin{equation}
 u_x(y)=4u_0\frac{y}{h}\left(1-\frac{y}{H}\right),\quad  \quad 0\leq y\leq H,
\label{eq:3.7}
\end{equation}
where $u_0=\bar{F}H^2/(8\rho_0\nu)$ is the maximum velocity, $H$ is the
channel height and $\bar{F}$ is the driving force. The Reynolds number
$Re=Hu_0/\nu$ is related to maximum velocity and pipe height.

\begin{figure}[h]
\subfigure[]{ \label{fig:poiseuille:a}
\includegraphics[scale=0.39]{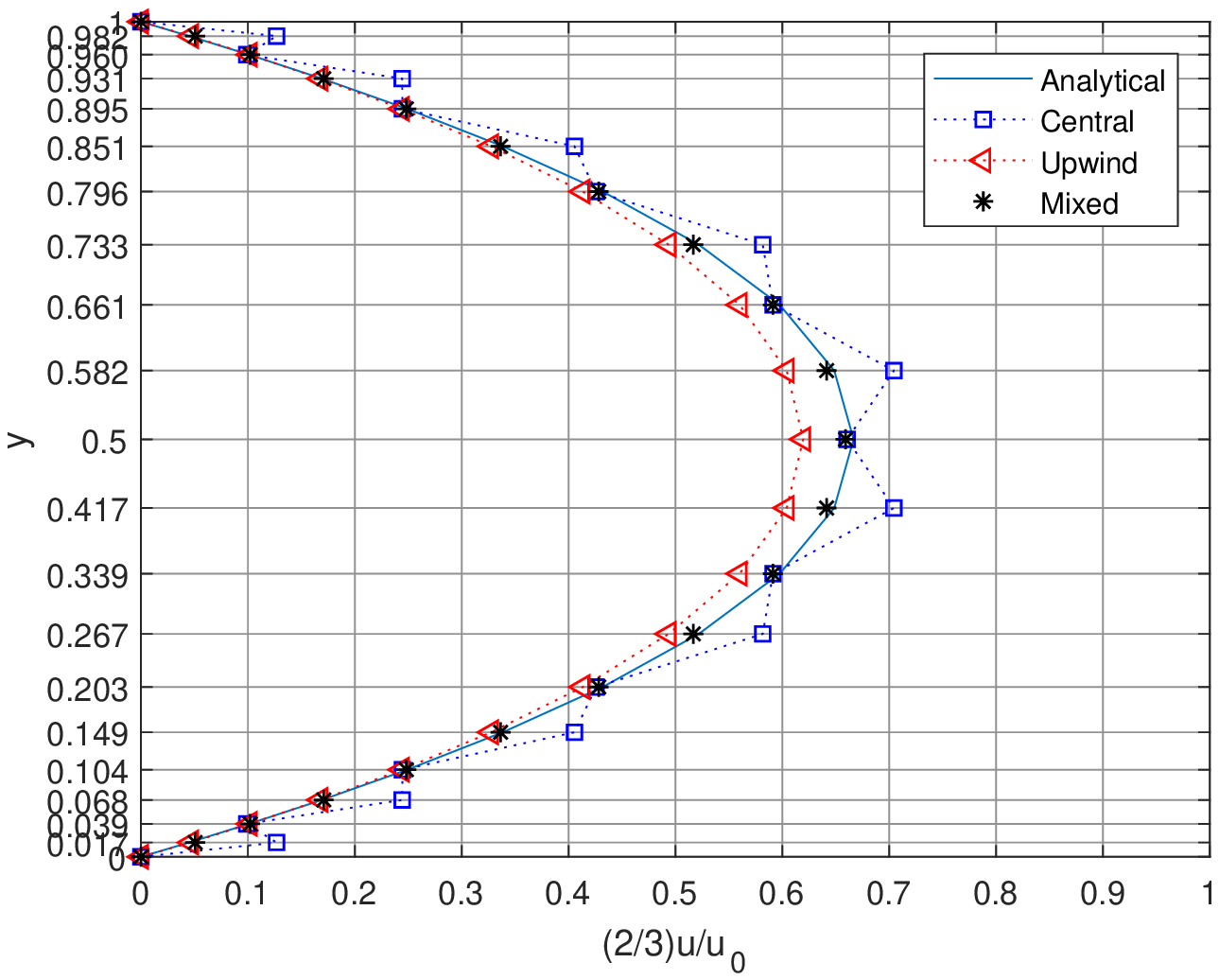}}
\subfigure[]{ \label{fig:poiseuille:b}
\includegraphics[scale=0.39]{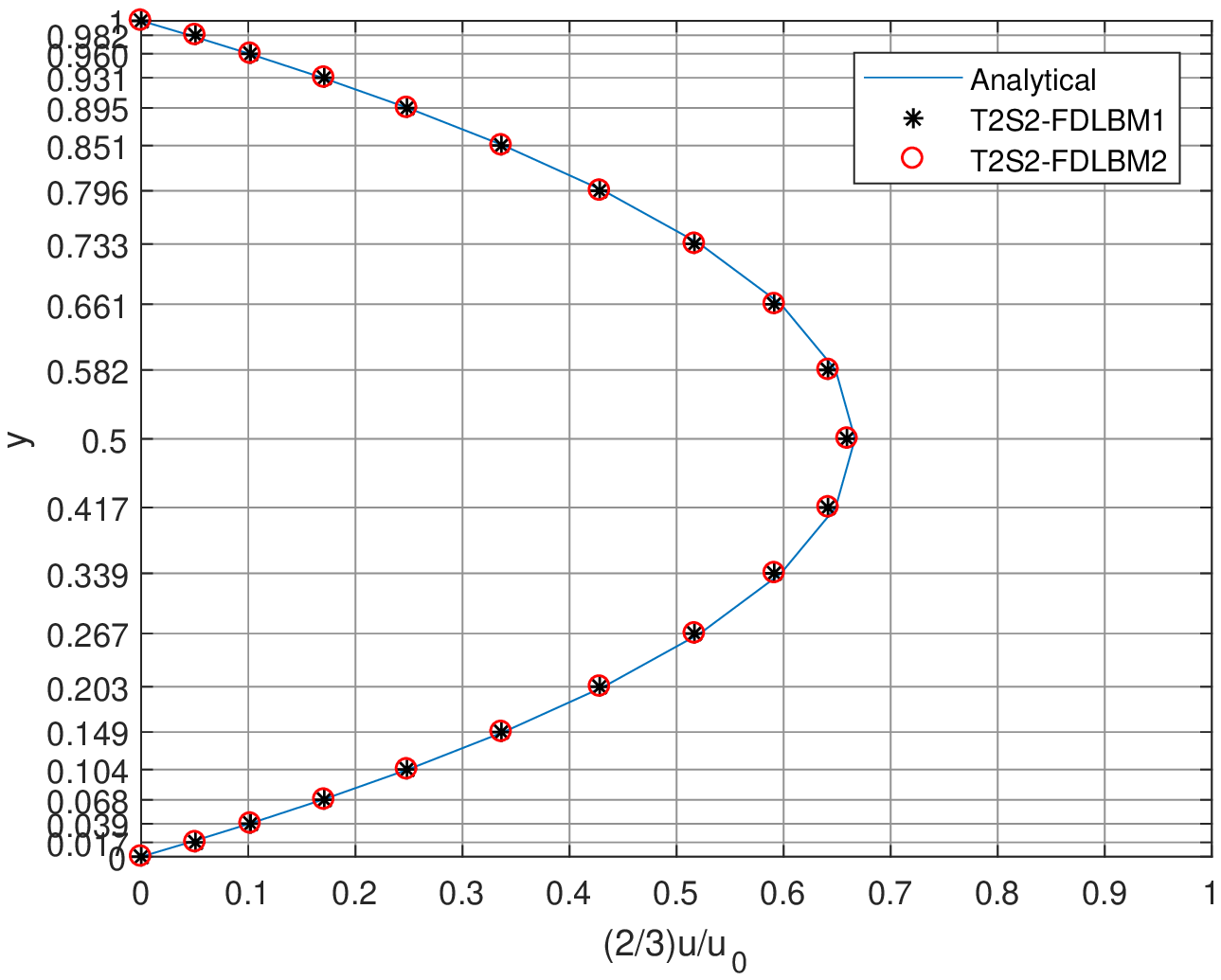}}
\caption{Distributions of non-uniform grids and the velocity along the centerline. } \label{fig:poiseuille}
\end{figure}

In our simulations, $L=H=1.0$ and $Re=10.0$. The periodic boundary condition
is used at the inlet and outlet of the channel, and the nonequilibrium
extrapolation scheme is applied to treat the nonslip boundary condition at
both top and bottom walls, which can be given by
\begin{equation}
 \tilde{f}(\bm{x}_b,\bm{\xi},t)=f_i^{eq}(\bm{x}_b,\bm{\xi}_i,t)+[\tilde{f}_i(\bm{x}_j,\bm{\xi}_i,t)-f_i^{eq}(\bm{x}_j,\bm{\xi}_i,t)].
\label{eq:3.10}
\end{equation}
Initially, the density $\rho=1.0$, $u=v=0.0$, the distribution function is
initialized by Eq.~\eqref{eq:3.2}. For this problem, the non-uniform gird is
applied to improve the computational efficiency, which is given by the
following transformation,
\begin{equation}
 x=\zeta,\quad y=\frac{1}{2a}[a+\tanh(c\mu)],
\label{eq:3.9}
\end{equation}
where $c$ is used to adjust the distribution of the grid and $a=\tanh(c)$.
$(\zeta,\mu)$ is the point of grid specified by $\zeta_i=i/N_x$ and
$\mu_j=(2j-N_y)/N_y$, where $i=0,1,...,N_x$ and $j=0,1,...,N_y$. In our
simulations, we set $N_x\times N_y=10\times 20$ and $c=1.5$.
Fig.~\ref{fig:poiseuille} shows the distribution of non-uniform grids. The
driving force $\bar{F}$ is chosen to be 0.01 to keep the maximum velocity
$u_0$ small, the time step can be set as $\Delta t=0.1\times y_1$, where
$y_1$ represents the height from the bottom to the first layer of the grid.
\begin{table}[!htbp]
\caption{Errors for the different difference schemes Poiseuille flow.}
  \centering
  \vspace{1ex}
\resizebox{\textwidth}{!}{
\begin{tabular}{clccccccc}
\hline\hline
\            & $\qquad$  & T1S2-FDLBM  & $\qquad$ &   T2S2-FDLBM1   & $\qquad$ &  T1S2-FDLBM   & $\qquad$ &  T2S2-FDLBM1         \\
\            & $\qquad$  & center error        & $\qquad$ &   center error        & $\qquad$ &  $GRE$         & $\qquad$ &  $GRE$               \\
\hline
\  up-wind & $\qquad$  & $3.719\times 10^{-5}$       & $\qquad$ &   $3.287\times 10^{-5}$         & $\qquad$ &  $4.502\times 10^{-6}$        & $\qquad$ &  $1.291\times 10^{-6}$              \\
\  central & $\qquad$  & $9.212\times 10^{-6}$     & $\qquad$ &   $5.098\times 10^{-6}$       & $\qquad$ &  $9.615\times 10^{-6}$      & $\qquad$ &  $2.705\times 10^{-6}$            \\
\  mixed  & $\qquad$  & $4.479\times 10^{-6}$     & $\qquad$ &   $2.641\times 10^{-6}$       & $\qquad$ &  $5.708\times 10^{-6}$      & $\qquad$ &  $2.538\times 10^{-6}$            \\
\hline\hline
\end{tabular}
}\label{table:poiseuille_3scheme}
\end{table}
\begin{table}[!htbp]
\caption{$GRE$s and temporal accuracy orders of FDLBM with non-uniform grid.}
 \label{table:poiseuille_GRE}
  \centering
  \vspace{1ex}
\resizebox{\textwidth}{!}{
\begin{tabular}{ccccccccccccccc}
\hline\hline
\  Model    & $\quad$  & $\Delta t$      & $\quad$  & $0.2\times y1$        & $\quad$ &  $0.3\times y1$      & $\quad$  &   $0.4\times y1$        & $\quad$  & $0.5\times y1$       & $\quad$  &  $0.6\times y1$       & $\quad$ &  $0.7\times y1$          \\
\hline

\  T1S2-       & $\quad$  & E(u)  & $\quad$  & $1.09\times 10^{-5}$ & $\quad$ & $1.51\times 10^{-5}$ & $\quad$  &   $ 1.98\times 10^{-5}$ & $\quad$  & $2.29\times 10^{-5}$ & $\quad$  &  $-$ & $\quad$ & $-$     \\

\  FDLBM       & $\quad$  & order & $\quad$  & $--$                 & $\quad$ & $0.8004$             & $\quad$  & $0.9435$                & $\quad$  & $ 0.6549$            & $\quad$  & $-$              & $\quad$ &  $-$     \\

\  T2S2-       & $\quad$  & E(u)  & $\quad$  & $6.45\times 10^{-6}$ & $\quad$ & $1.35\times 10^{-5}$ & $\quad$  &   $ 2.39\times 10^{-5}$ & $\quad$  & $3.69\times 10^{-5}$ & $\quad$  &  $5.38\times 10^{-5}$ & $\quad$ & $7.35\times 10^{-5}$     \\

\  FDLBM1       & $\quad$  & order & $\quad$  & $--$                 & $\quad$ & $1.8235$             & $\quad$  & $1.9911$                & $\quad$  & $ 1.9434$            & $\quad$  & $2.0640$              & $\quad$ &  $2.0187$     \\
\  T2S2-       & $\quad$  & E(u)  & $\quad$  & $3.34\times 10^{-4}$ & $\quad$ & $7.62\times 10^{-4}$ & $\quad$  &   $ 1.38\times 10^{-3}$ & $\quad$  & $2.20\times 10^{-3}$ & $\quad$  &  $3.23\times 10^{-3}$ & $\quad$ & $4.52\times 10^{-3}$     \\

\  FDLBM2       & $\quad$  & order & $\quad$  & $--$                 & $\quad$ & $2.0337$             & $\quad$  & $2.0587$                & $\quad$  & $ 2.0889$            & $\quad$  & $2.1258$              & $\quad$ &  $2.1696$     \\
\hline\hline
\end{tabular}
}
\end{table}
\begin{figure}[h]
\centering
\includegraphics[scale=0.5]{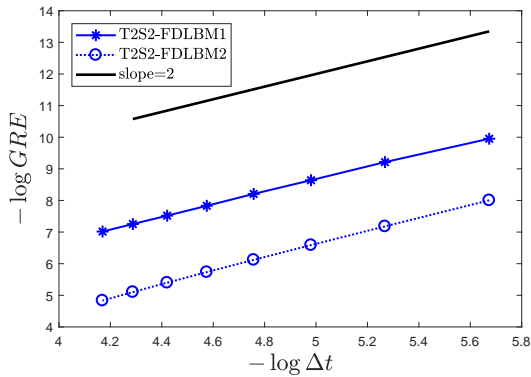}
\caption{The $GRE$ of T2S2-FDLBM1 with the non-uniform grid at different time steps.} \label{fig:Torder}
\end{figure}

Fig.~\ref{fig:poiseuille:a} displays the numerical results of T2S2-FDLBM1
with the up-wind, central and mixed difference scheme ($\eta=0.1$). The
results of T2S2-FDLBM1 and T2S2-FDLBM2 are also shown in
Fig.~\ref{fig:poiseuille:b}. From this figure, it can be observed that there
are some numerical oscillations in the central difference scheme, and the
phenomenon of numerical dissipation appears in the second-order upwind
difference scheme. In general, the result of mixed scheme is the most
accurate, which is similar to that of the T1S2-FDLBM \cite{guo2003explicit}.
However, it should be noted that the numerical oscillation of the central
difference scheme in T2S2-FDLBM1 is much smaller than T1S2-FDLBM
\cite{guo2003explicit}. This also indicates that T2S2-FDLBM1 is more stable.
In addition, as seen from Fig.~\ref{fig:poiseuille:b}, the results of
T2S2-FDLBM1 and T2S2-FDLBM2 with mixed difference scheme agree well with the
analytical solution when $CFL=0.1$. The error of velocity at the centerline
is given in Table~\ref{table:poiseuille_3scheme}. From this table, one can
find that the T2S2-FDLBM1 is more accurate than T1S2-FDLBM.
\begin{table}[!htbp]
\caption{$GRE$s of Poiseuille flow with different values of $CFL$ condition number.}
  \label{table:poiseuille_CFL}
  \centering
  \vspace{1ex}
\resizebox{\textwidth}{!}{
\begin{tabular}{ccccccccccccccccccccc}
\hline\hline
\   $CFL$    & $\quad$  & $0.1$         & $\quad$ &   $0.2$          & $\quad$ &  $0.3$           & $\quad$  &  $0.4 $
             & $\quad$  &  $0.5 $       & $\quad$ &  $0.6 $          & $\quad$ &  $0.7 $          & $\quad$  &  $0.8 $
             & $\quad$  &  $0.9 $ \\
\hline
\ T1S2-FDLBM $GRE$$(\times 10^{-4})$ & $\quad$  &  $0.0961$     & $\quad$ &   $0.1092$       & $\quad$ &  $0.1510$             & $\quad$  &  $0.1981$
             & $\quad$  &  $0.2293$         & $\quad$ &  $- $            & $\quad$ &  $- $            & $\quad$  &  $- $
             & $\quad$  &  $- $ \\
\ T2S2-FDLBM1 $GRE$$(\times 10^{-4})$ & $\quad$  &  $0.0254$     & $\quad$ &   $0.0645$       & $\quad$ &  $0.1350$             & $\quad$  &  $0.2394$
             & $\quad$  &  $0.3693$         & $\quad$ &  $0.5381$            & $\quad$ &  $0.7345$            & $\quad$  &  $0.9582$
             & $\quad$  &  $1.2137$ \\
\ T2S2-FDLBM2 $GRE$$(\times 10^{-4})$ & $\quad$  &  $0.0254$     & $\quad$ &   $3.3396$       & $\quad$ &  $7.6177$             & $\quad$  &  $13.7733$
             & $\quad$  &  $21.9519$         & $\quad$ &  $32.3438$            & $\quad$ &  $45.1894$            & $\quad$  &  $60.7891$
             & $\quad$  &  $79.5211$ \\
\hline\hline
\end{tabular}
}
\end{table}

\renewcommand{\multirowsetup}{\centering}
\begin{table}[!htbp]
\centering
\caption{A comparison of four different methods for Poiseuille flow at $t=20s$.}
\label{table:poiseuille_comparison}
 \centering
\vspace{1ex}
\resizebox{\textwidth}{!}{
\begin{tabular}{m{6em}<{\centering}m{1em}m{6em}<{\centering}m{6em}<{\centering}m{1em}m{6em}<{\centering}m{6em}<{\centering}m{1em}m{6em}<{\centering}m{6em}<{\centering}
m{1em}m{6em}<{\centering}m{6em}<{\centering}}
\hline\hline
\  model  &  & \multicolumn{2}{c}{T2S2-FDLBM1}&  & \multicolumn{2}{c}{T2S2-FDLBM2}   &     &    \multicolumn{2}{c}{T1S2-FDLBM}            &  &  \multicolumn{2}{c}{SLBM}               \\
\cline{3-4}\cline{6-7}\cline{9-10}\cline{12-13}
\  grid & & $10\times 20$ & $10\times 20$ & & $10\times 20$ & $10\times 20$ & & $10\times 20$ & $10\times 20$ & & $80\times 80$ & $20\times 20$   \\
\  $CFL$ & & $0.1$ & $0.9$ & & $0.1$ & $0.9$ & & $0.1$ & $0.5$ & & -- & --    \\
\ $ \Delta t$ & & $0.0017$ & $0.0155$ & & $0.0017$ & $0.0155$ & & $0.0017$ & $0.0086$ & & $0.0125$ & $0.0500$  \\
\  iterative times & & $11643$ & $1293$ & & $11643$ & $1293$ & & $11643$ & $2328$ & & $1600$ & $400$  \\
\cline{3-4}\cline{6-7}\cline{9-10}\cline{12-13}
\  CPU time & & $6.1400$ & $0.7500$ & & $7.6690$ & $0.9090$ & & $3.8590$ & $0.7970$ & & $6.3750$ & $0.1720$  \\
\  ratio & & $8.1867$ & $1.0000$ & & $10.2253$ & $1.2120$ & & $5.1453$ & $1.0627$ & & $8.5000$ & $0.2293$  \\
\cline{3-4}\cline{6-7}\cline{9-10}\cline{12-13}
\  $GRE$ & & $1.3618\times 10^{-5}$ & $1.2155\times 10^{-4}$ & & $8.2600\times 10^{-5}$ & $7.9521\times 10^{-3}$ & & $2.7805\times 10^{-5}$ & $1.3623\times 10^{-4}$ & & $1.7095\times 10^{-4}$ & $6.8185\times 10^{-4}$  \\
\  ratio & & $0.1120$ & $1.0000$ & & $0.6796$ & $65.4225$ & & $0.2288$ & $ 1.1208$ & & $1.4064$ & $5.6096$  \\
\hline\hline

\end{tabular}
}
\end{table}

To test the convergence order of two T2S2-FDLBMs in time, the $GRE$s at
different time steps are calculated in Table~\ref{table:poiseuille_GRE}. It
can be seen that the T1S2-FDLBM is only first-order accurate in time, while
the T2S2-FDLBM1 and T2S2-FDLBM2 have a second-order convergence rate, which
is also consistent with the theoretical analysis. In addition, we also
tested the effect of $CFL$ condition number, and presented the results in
Table~\ref{table:poiseuille_CFL}. From this table, it can be found that the
maximum values of $CFL$ condition number in T2S2-FDLBM1 and T2S2-FDLBM2 can
reach to 0.9, while it is only about $0.5$ in T1S2-FDLBM. The $GRE$s of
T2S2-FDLBM2 are larger than T2S2-FDLBM1 as $CFL$ condition number increases.
In Table~\ref{table:poiseuille_comparison}, we presented a comparison of the
computational efficiency of four methods. Under the condition of similar
error, the CPU time of SLBM is $8.5$ times as long as T2S2-FDLBM. While
under the condition of similar CPU time, the $GRE$ of T1S2-FDLBM is a little
larger than that of T2S2-FDLBM1, and the $GRE$ of T2S2-FDLBM2 is $65.4$
times as large as T2S2-FDLBM1. Therefore, compared with other three methods,
T2S2-FDLBM1 is more efficient.\\

\textbf{C. The lid-driven cavity flow}

As a classic problem, the lid-driven cavity flow is also used to test
T2S2-FDLBM1. The lid-driven cavity flow is driven by a constant velocity of
the top wall, and the other three solid walls remain stationary. To obtain
accurate results, it is necessary to refine the grid at the four corners,
this is because the flow phenomenon at the four corners are very complex
\cite{Zhen_Hua_2006}.

T2S2-FDLBM1 is applied to simulate the lid-driven flow in a square cavity.
The height of the square cavity is set to be $1.0$. The top wall moves
horizontally from left to right with a constant velocity $u_0=0.1$. The
initial density and velocity are chosen to be $\rho=1.0$ and
$\textbf{u}=\textbf{0}$. The boundary conditions are treated by the
non-equilibrium extrapolation scheme. The non-uniform is also applied for
this problem,
\begin{equation}
 x=\frac{1}{2a}[a+\tanh(c\zeta)],\quad y=\frac{1}{2a}[a+\tanh(c\mu)],
\label{eq:3.11}
\end{equation}
where $c=1.5$ and $a=\tanh(c)$. $(\zeta,\mu)$ is the point of grid set by
$\zeta_i=i/N_x$ and $\mu_j=j/N_y$, where $i=0,1,...,N_x$ and
$j=0,1,...,N_y$. In our simulations, $N_x\times N_y=64\times 64$ for
$Re=400$ and $1000$, $N_x\times N_y=128\times 128$ for$Re=3200$ and $5000$,
the time step is set to be $\Delta t=0.1\times y_1$. In order to eliminate
the numerical dissipation, the parameter $\eta$ is set to be 0.1 for
$Re=400$ and $1000$, and 0.05 for$Re=3200$ and $5000$. The $GRE$ of
lid-driven cavity flow can be defined as
\begin{equation}
\ E(u) = \frac{\sqrt{\sum_{i,j} |u_{i,j}(t_n)-u_{i,j}(t_{n-1})|^2}}{\sqrt{\sum_{i,j} |u_{i,j}(t_{n})|^2}}.
\label{eq:GRE}
\end{equation}

\begin{figure}[h]
\subfigure[$Re=400$]{ \label{fig:Re_400}
\includegraphics[scale=0.25]{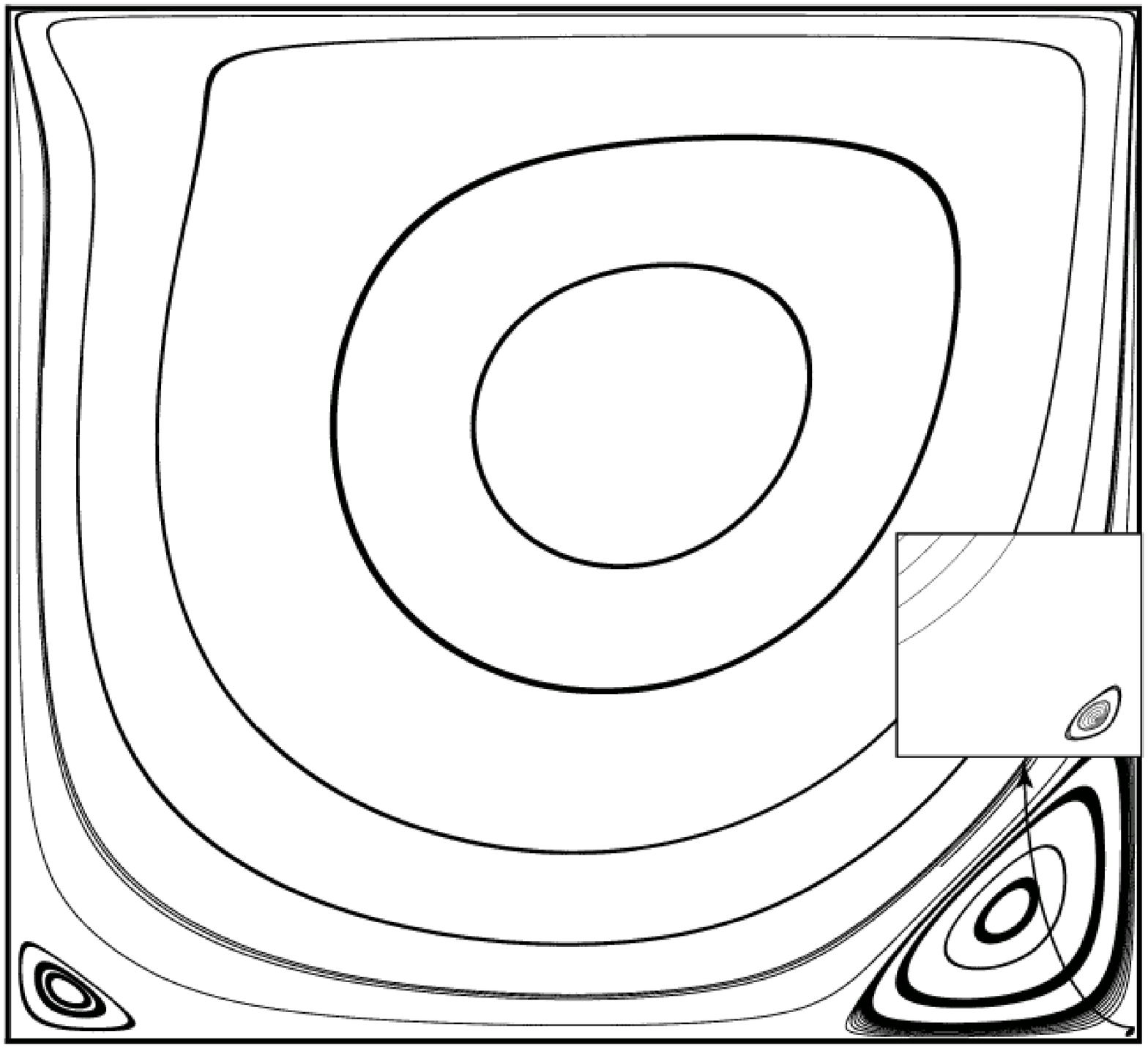}}
\subfigure[$Re=1000$]{ \label{fig:Re_1000}
\includegraphics[scale=0.25]{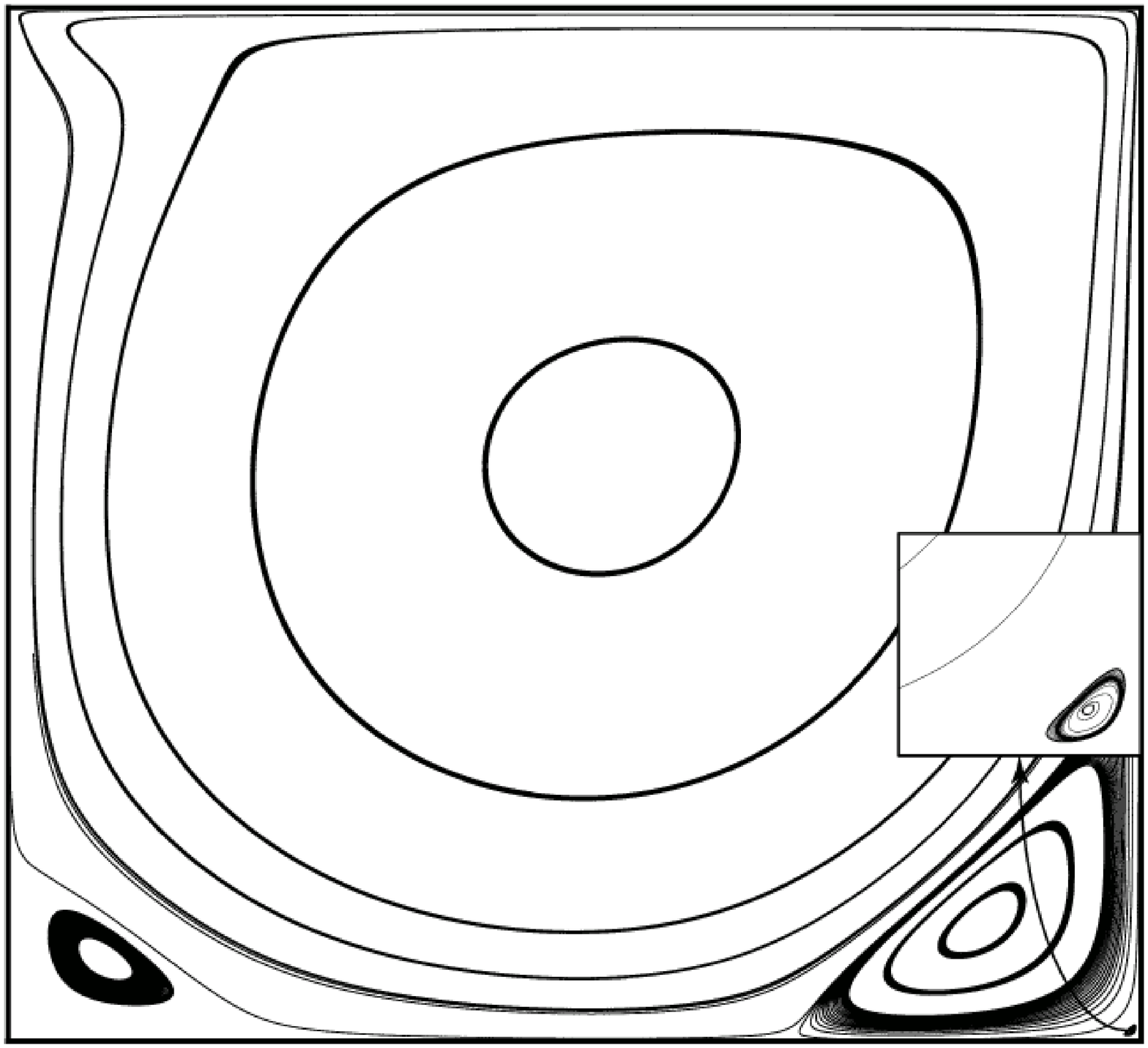}}
\subfigure[$Re=3200$]{ \label{fig:Re_3200}
\includegraphics[scale=0.25]{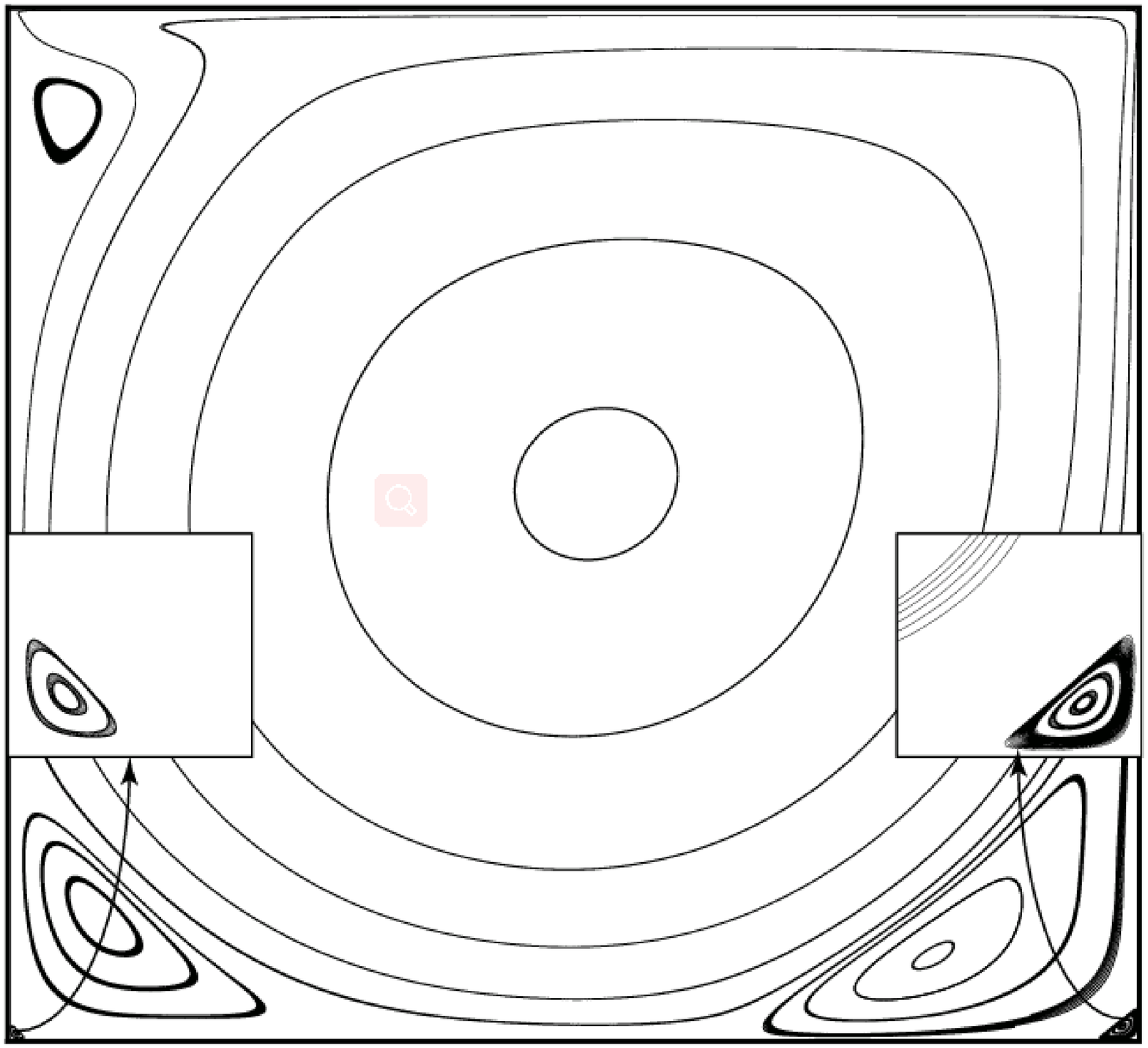}}
\subfigure[$Re=5000$]{ \label{fig:Re_5000}
\includegraphics[scale=0.25]{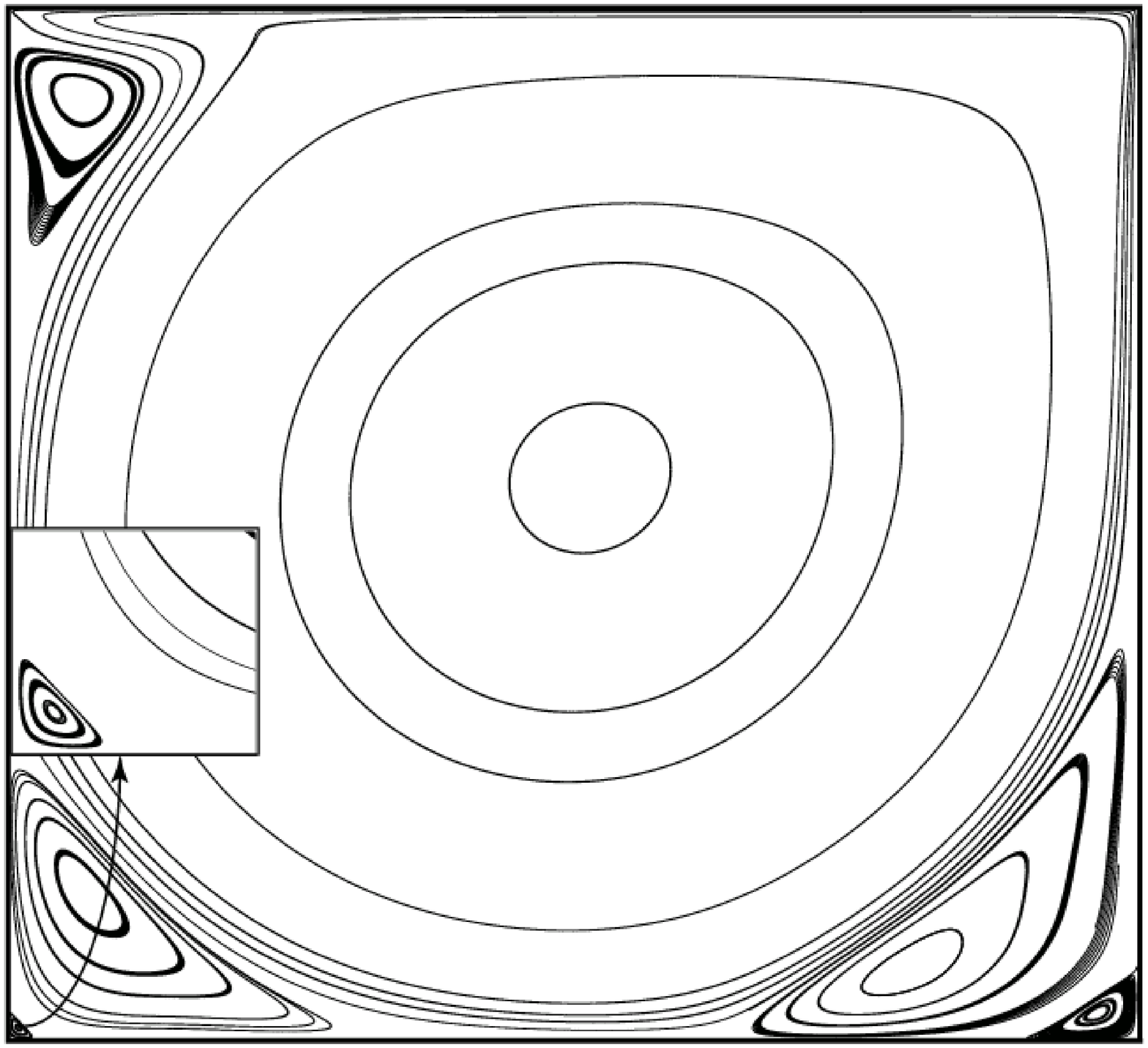}}
\caption{Streamlines of lid-driven cavity flow with different Reynolds.}
\label{fig:Lid-driven flow}
\end{figure}

\begin{figure}[h]
\subfigure[]{ \label{fig:Lid-u}
\includegraphics[scale=0.39]{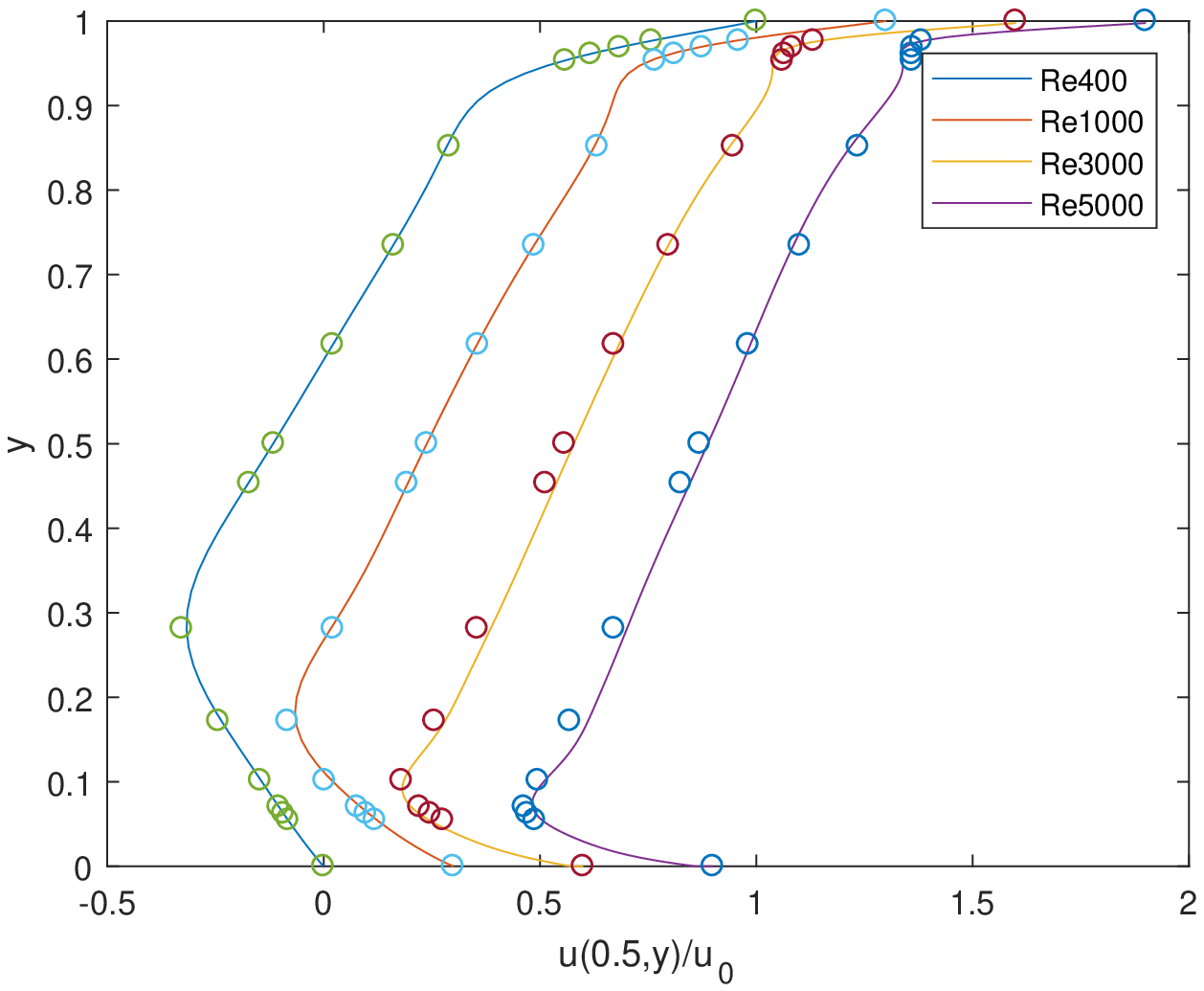}}
\subfigure[]{ \label{fig:Lid-v}
\includegraphics[scale=0.39]{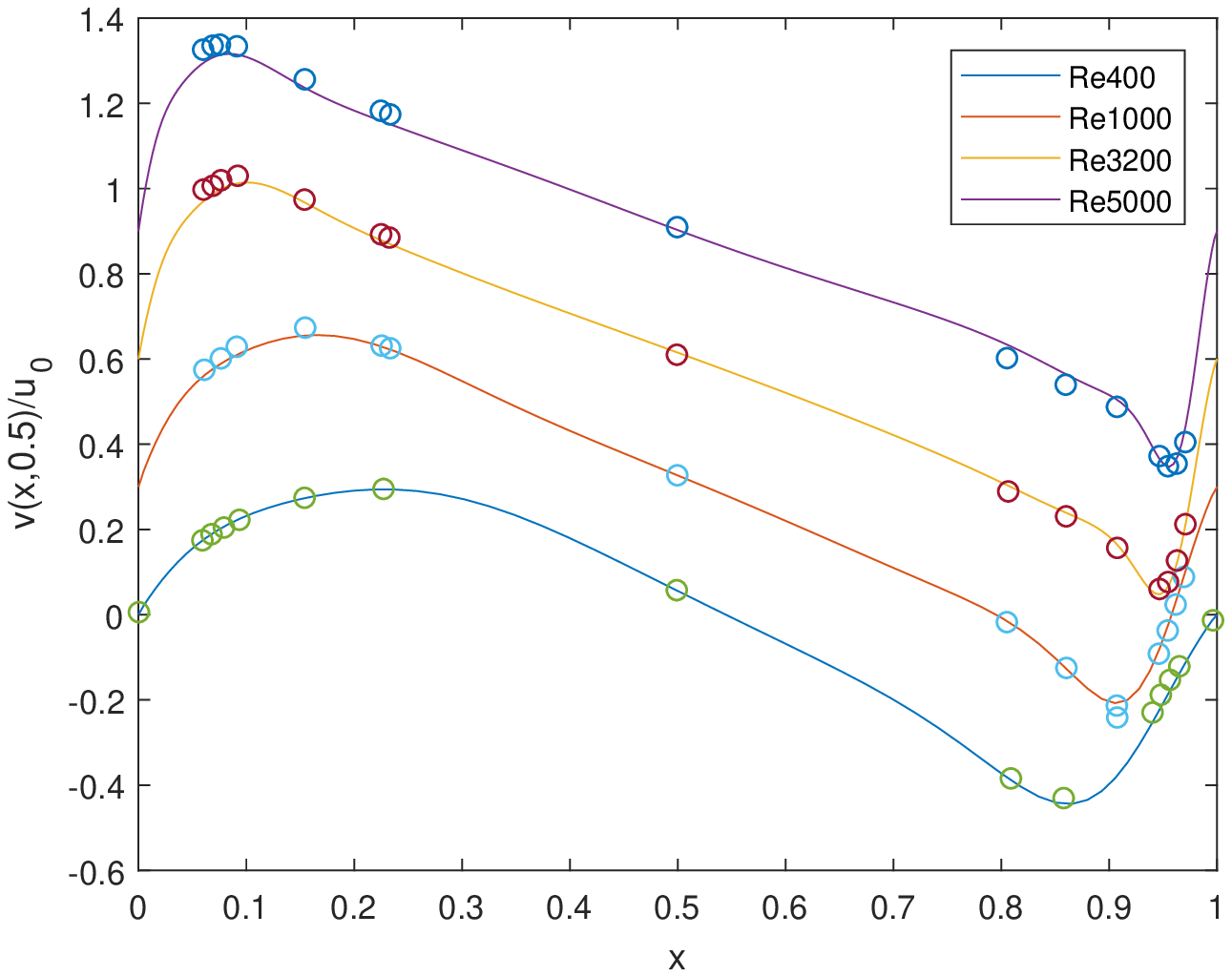}}
\caption{Velocity profiles along the centerline at different $Re$, (o) is reference date. [(a) $Re$=400, 1000, 3200, 5000 from left to right; (b) $Re$=400, 1000, 3200, 5000 from bottom to top.]}
\label{fig:Lid-uv}
\end{figure}

\renewcommand{\multirowsetup}{\centering}
\begin{table}[!htbp]
\caption{$GRE$s of lid-driven cavity flow with different values of $CFL$ condition number.}
 \label{table:lid_CFL}
  \centering
  \vspace{1ex}
\resizebox{\textwidth}{!}{
\begin{tabular}{ccccccccccccccccccccc}
\hline\hline
\   $CFL$    & $\quad$  & $0.1$         & $\quad$ &   $0.2$          & $\quad$ &  $0.3$           & $\quad$  &  $0.4 $
             & $\quad$  &  $0.5 $       & $\quad$ &  $0.6 $          & $\quad$ &  $0.7 $          & $\quad$  &  $0.8 $
             & $\quad$  &  $0.9 $ \\
\hline
\ T1S2-FDLBM $GRE$ & $\quad$  &  $0.5816$     & $\quad$ &   $0.4934$       & $\quad$ &  $-$             & $\quad$  &  $- $
             & $\quad$  &  $- $         & $\quad$ &  $- $            & $\quad$ &  $- $            & $\quad$  &  $- $
             & $\quad$  &  $- $ \\
\ T2S2-FDLBM1 $GRE$ & $\quad$  &  $0.5725$     & $\quad$ &   $0.4858$       & $\quad$ &  $0.4681$             & $\quad$  &  $0.4609 $
             & $\quad$  &  $0.4565 $         & $\quad$ &  $0.4542 $            & $\quad$ &  $0.4524$            & $\quad$  &  $0.4517$
             & $\quad$  &  $0.4510 $ \\
\hline\hline
\end{tabular}
}
\end{table}

\renewcommand{\multirowsetup}{\centering}
\begin{table}[!htbp]
  \vspace{1ex}
\caption{The vortices location of lid-driven cavity flow.} \label{table:lid
location}
\centering
 \vspace{1ex} \resizebox{\textwidth}{!}{
\begin{tabular}{m{5em}<{\centering}m{1em}m{8em}<{\centering}m{1em}m{6em}<{\centering}m{6em}<{\centering}m{1em}m{6em}<{\centering}m{6em}<{\centering}m{1em}m{6em}<{\centering}m{6em}<{\centering}}
\hline\hline
\  &  & &  & \multicolumn{2}{c}{Primary Vortex}   &     &    \multicolumn{2}{c}{Left Lower Vortex}            &  &  \multicolumn{2}{c}{Right Lower Vortex}               \\
\cline{5-6}\cline{8-9}\cline{11-12}
 \  &  & &  & $X$   & $Y$   &     &    $X$   & $Y$   &  &  $X$   & $Y$     \\
 \hline
  \multirow{2}{5em}{$Re=1000$} & &SLBM \cite{ghia1982high} & & $0.5313$ & $0.5625$ & & $0.0859$ & $0.0781$ & & $0.8594$ & $0.1094$ \\
   & &T2S2-FDLBM1 & & $0.5330$ & $0.5670$ & & $0.0850$ & $0.0796$ & & $0.8600$ & $0.1136$ \\
 \hline
  \multirow{2}{5em}{$Re=5000$} & &SLBM \cite{hou1995simulation} & & $0.5176$ & $0.5373$ & & $0.0784$ & $0.1373$ & & $0.8078$ & $0.0745$ \\
   & &T2S2-FDLBM1 & & $0.5162$ & $0.5460$ & & $0.0704$ & $0.1442$ & & $0.7990$ & $0.0699$ \\
\hline\hline

\end{tabular}
}
\end{table}

Fig.~\ref{fig:Lid-driven flow} shows streamline of the lid-drive flow at
different values of Reynolds number. It can be observed that four vortices
appear in the cavity when $Re\leq 1000$: a primary vortex at the center of
the cavity, a pair of secondary vortices at the lower left and lower right
corners, a third level vortex at the lower right corner. When $Re$ is up to
$3200$ or $5000$, a third secondary vortex appears in the upper left corner.
As $Re$ increases, the center of the primary vortex approaches the center of
the cavity. Compared with the results of SLBM \cite{ghia1982high},
T2S2-FDLBM1 can capture more flow details even for $N_x\times N_y=64\times
64$. Fig.~\ref{fig:Lid-uv} displays the velocity $u$ and $v$ along the
centerline of the cavity. It can be found that the results are in good
agreement with the previous work
\cite{ghia1982high,schreiber1983driven,vanka1986block}. In
Table~\ref{table:lid location}, the locations of the vortices are also
consistent with the available results \cite{ghia1982high,
hou1995simulation}. From Tab.~\ref{table:lid_CFL}, it is observed that the
range of the $CFL$ condition number in T2S2-FDLBM1 is larger than that in
SLBM. Besides, the stability of SLBM, T1S2-FDLBM and T2S2-FDLBM1 are also
tested with this example. Under a small grid size ($64\times 64$), the SLBM
will be divergent when $Re>8600$, but T1S2-FDLBM and T2S2-FDLBM1 can work
well even for $Re\geq 20000$.

\section{\label{sec:level4}Conclusions}
In this work, a class of T2S2-FDLBM with a second-order accuracy in time and
space is proposed based on T1S2-FDLBM presented by Guo et al.
\cite{guo2003explicit}. In this method, a simplified TFTD method is applied
for time discretization, and a mixed difference scheme is used for space
discretization. It is also shown that the T1S2-FDLBM is just a special case
of the T2S2-FDLBM. Through the stability analysis, two specific T2S2-FDLBMs
are determined. We also performed some simulation to test two T2S2-FDLBMs,
and the results are in good agreement with analytical solutions or some
previous work. In addition, it is shown that the T2S2-FDLBM1 has a
second-order accuracy both in time and space, and the non-uniform grid is
also applied to improve computational efficiency. Compared with the SLBM,
T1S2-FDLBM and T2S2-FDLBM2, T2S2-FDLBM1 can give more accurate results, and
is also more efficient. On the other hand, the CFL condition number in two
T2S2-FDLBMs can be changed in a larger range, this feature can be also used
to remove the limitation of time step in T1S2-FDLBM. Finally, T2S2-FDLBM1 is
more stable, and the numerical oscillations can be reduced effectively.
Moreover, T2S2-FDLBM can be also extended to nonlinear convection-diffusion
equation, which would be discussed in a future work.

\section*{Acknowledgements}
This work is supported by the National Natural Science Foundation of China
(Grants No. 51836003 and No. 51576079), and the National Key Research and
Development Program of China (Grant No. 2017YFE0100100).

\section*{References}

\bibliography{elsarticle-template}

\end{document}